\documentclass[12pt,a4paper]{article}

\usepackage{fixltx2e}

\usepackage[font=small]{caption}
\usepackage[text={450pt,650pt},headheight={14.5pt},centering]{geometry}

\usepackage{lmodern}
\usepackage{graphicx}
\usepackage{subfig}
\usepackage{booktabs}
\usepackage{multirow}
\usepackage{dsfont}

\usepackage{cite}
\usepackage[bulletsep]{collref}

\usepackage{tikz}
\usetikzlibrary{plotmarks,calc,decorations.markings,decorations.pathmorphing,patterns}

\makeatletter
\pgfarrowsdeclare{der}{der}% Hack to make an arrow that is centered on the "tip". Based on the default tikz arrow. I have only checked that this looks OK with thick lines.
{
  \pgfutil@tempdima=0pt%
  \pgfutil@tempdimb=-0.28pt%
  \advance\pgfutil@tempdimb by-0.3\pgflinewidth%
  \pgfarrowsleftextend{+\pgfutil@tempdima}
  \pgfarrowsrightextend{+1.125\pgfutil@tempdimb}
}
{
  \pgfutil@tempdima=0.28pt%
  \advance\pgfutil@tempdima by.3\pgflinewidth%
  \pgfsetlinewidth{0.8\pgflinewidth}
  \pgfsetdash{}{+0pt}
  \pgfsetroundcap
  \pgfsetroundjoin
  \pgfpathmoveto{\pgfqpoint{-3\pgfutil@tempdima}{4\pgfutil@tempdima}}
  \pgfpathcurveto
  {\pgfqpoint{-2.75\pgfutil@tempdima}{2.5\pgfutil@tempdima}}
  {\pgfqpoint{0pt}{0.25\pgfutil@tempdima}}
  {\pgfqpoint{0.75\pgfutil@tempdima}{0pt}}
  \pgfpathcurveto
  {\pgfqpoint{0pt}{-0.25\pgfutil@tempdima}}
  {\pgfqpoint{-2.75\pgfutil@tempdima}{-2.5\pgfutil@tempdima}}
  {\pgfqpoint{-3\pgfutil@tempdima}{-4\pgfutil@tempdima}}
  \pgfusepathqstroke
}
\makeatother

\tikzset{%
  scalar/.style={thick,double},
  fermion/.style={thick,double,dashed},
  gluon/.style={thick,double,decorate,decoration={snake,amplitude=.25mm,segment length=1.25mm,post length=0.2mm,pre length=0.2mm}},
  hyp scalar/.style={thick},
  hyp fermion/.style={thick,dashed},
  adj fermion/.style={thick,double,dash pattern=on 0.5mm off 0.5mm},
  arrow/.style 2 args={decoration={markings,mark=at position #1 with {\arrow[#2]{latex}}}},
  derivative/.style={decoration={markings,mark=at position #1 with {\arrow[scale=1.25]{der}}}}
}

\newcommand{\sL}{\mbox{\tiny L}}
\newcommand{\sR}{\mbox{\tiny R}}

\usepackage{amsmath,amssymb}
\usepackage{mathtools}
\usepackage{slashed}
\usepackage{braket}

\usepackage{hyperref}

\usepackage{xspace}

\usepackage{accents}

\numberwithin{equation}{section}

\makeatletter
 \let\old@startsection=\@startsection
 \let\oldl@section=\l@section
 \renewcommand{\@startsection}[6]{\old@startsection{#1}{#2}{#3}{#4}{#5}{#6\mathversion{bold}}}
 \renewcommand{\l@section}[2]{\oldl@section{\mathversion{bold}#1}{#2}}
\makeatother

\DeclareMathOperator{\tr}{tr}

\DeclareMathOperator{\Sym}{Sym}

\def\XXint#1#2#3{{\setbox0=\hbox{$#1{#2#3}{\int}$}
    \vcenter{\hbox{$#2#3$}}\kern-.5\wd0}}

\newcommand{\AdS}{\textup{AdS}}
\newcommand{\CFT}{\text{CFT}}
\newcommand{\Sphere}{\mathrm{S}}
\newcommand{\Torus}{\mathrm{T}}

\newcommand{\comm}[2]{[#1,#2]}
\newcommand{\acomm}[2]{\{#1,#2\}}

\newcommand{\alg}[1]{\mathrm{#1}}
\newcommand{\grp}[1]{\mathrm{#1}}

\newcommand{\algSU}{\alg{su}}
\newcommand{\grpSU}{\grp{SU}}

\newcommand{\algPU}{\alg{pu}}

\newcommand{\algSO}{\alg{so}}
\newcommand{\grpSO}{\grp{SO}}

\newcommand{\algU}{\alg{u}}
\newcommand{\grpU}{\grp{U}}

\newcommand{\algPSU}{\alg{psu}}

\newcommand{\grpUSp}{\grp{USp}}

\newcommand{\gen}[1]{\mathbf{#1}}

\newcommand{\rep}[1]{\mathbf{#1}}

\newcommand{\superN}{\mathcal{N}}

\newcommand{\ie}{\textit{i.e.}\xspace}
\newcommand{\eg}{\textit{e.g.}\xspace}

\begin{document}

\thispagestyle{empty}

\begin{flushright}\footnotesize\ttfamily
Imperial-TP-OOS-2014-05\\
HU-Mathematik-2014-33\\
HU-EP-14/49
\end{flushright}
\vspace{1em}

\begin{center}
\textbf{\Large\mathversion{bold} Integrability and the Conformal Field Theory \\ of the Higgs branch}

\vspace{2em}

\textrm{\large Olof Ohlsson Sax${}^1$, Alessandro Sfondrini${}^{2}$
%\\
and Bogdan Stefa\'nski, jr.${}^3$ } 

\vspace{4em}

\begingroup\itshape

1. The Blackett Laboratory, Imperial College,\\  SW7 2AZ, London,
U.K.\\[0.2cm]

2. Institut f\"ur Mathematik und Institut f\"ur Physik, Humboldt-Universit\"at zu Berlin\\
IRIS Geb\"aude, Zum Grossen Windkanal 6, 12489 Berlin, Germany
\\[0.2cm]

3. Centre for Mathematical Science, City University London,\\ Northampton Square, EC1V 0HB, London, U.K.\\[0.2cm]
\par\endgroup

\vspace{1em}

%\texttt{o.olsson-sax@imperial.ac.uk, \\ Alessandro.Sfondrini@physik.hu-berlin.de, \\ Bogdan.Stefanski.1@city.ac.uk}

%%%%%%%%

\end{center}

\vspace{3em}

\begin{abstract}\noindent
In the context of the $\AdS_3/\CFT_2$ correspondence, we investigate the Higgs branch $\CFT_2$. Witten showed that states localised near the small instanton singularity can be described in terms of vector multiplet variables. This theory has a planar, weak-coupling limit, in which anomalous dimensions of single-trace composite operators can be calculated. At one loop, the calculation reduces to finding the spectrum of a spin-chain with 
nearest-neighbour interactions. This $\CFT_2$ spin-chain matches precisely the one that was previously found as the weak-coupling limit of the integrable system describing the $\AdS_3$ side of the duality.
We compute the one-loop dilatation operator in a non-trivial compact subsector and show that it corresponds to an integrable spin-chain Hamiltonian. This provides the first direct evidence of integrability on the $\CFT_2$ side of the correspondence. 
\end{abstract}

%%%%%%%%%%%%%%%%%%%%%%%%%%%%%%%%%%%%%%%%%%%%%%%%%%%%%%%%%%%%%%%%%%%%%%%%%%%
\newpage
\tableofcontents	

\section{Introduction}

The $\AdS/\CFT$ correspondence~\cite{Maldacena:1997re,Witten:1998qj,Gubser:1998bc} can be understood quantitatively through
integrability methods in certain classes of $\AdS_5/\CFT_4$ and $\AdS_4/\CFT_3$ dual pairs. Reviews of this approach and references to the literature can be found in~\cite{Arutyunov:2009ga, Beisert:2010jr}. Superstrings on $\AdS_3$ backgrounds with 16 supersymmetries are known to be classically integrable~\cite{Babichenko:2009dk,Sundin:2012gc,Cagnazzo:2012se}, and so one may wonder whether  integrability underlies the  $\AdS_3/\CFT_2$ correspondence.%
\footnote{For a review of the generalities of the $\AdS_3/\CFT_2$ correspondence see~\cite{David:2002wn}. Some aspects of the integrability approach to $\AdS_3/\CFT_2$ have been recently reviewed in~\cite{Sfondrini:2014via}.}
  Recently, we have found the complete non-perturbative world-sheet integrable S matrix of  string theory on such $\AdS_3$ backgrounds, including the hitherto unaccounted for \emph{massless} modes~\cite{Borsato:2014exa,Borsato:2014hja, Lloyd:2014bsa,bossstoappear}.\footnote{For some early work on integrability in the context of $\AdS_3/\CFT_2$ see also~\cite{David:2008yk,David:2010yg}.}  It appears now very likely that the $\AdS_3/\CFT_2$ correspondence with 16 supersymmetries is described, in the planar limit~\cite{'tHooft:1973jz}, through a holographic quantum integrable system. 

As a consequence, there exists an all-loop Bethe Ansatz (BA) whose solutions determine the closed-string spectrum in these backgrounds~\cite{Babichenko:2009dk,
OhlssonSax:2011ms,Borsato:2012ud, Borsato:2012ss,Borsato:2013qpa}.\footnote{The original conjecture for the BA~\cite{Babichenko:2009dk,OhlssonSax:2011ms} was put forward for \emph{massive} modes only. Later, it was found that the BA had to be modified when the non-perturbative worldsheet S matrix was constructed~\cite{Borsato:2012ud,Borsato:2012ss,Borsato:2013qpa}.} It is known that this BA, in the weak-coupling limit, encodes the energy spectrum of an integrable spin chain with local interactions~\cite{OhlssonSax:2011ms, Sax:2012jv, Borsato:2013qpa}. Based on experience with higher-dimensional holographic integrability, this spin-chain is believed to describe the large-curvature or small 't Hooft-coupling limit  of a discretised string world-sheet.

In higher-dimensional examples of integrable holography the local integrable spin-chain 
emerges in perturbative gauge theory calculations of anomalous dimensions of single-trace operators, as first shown in~\cite{Minahan:2002ve}. The relation between the spin-chain and perturbative gauge theory gives strong evidence of integrability of the dual pair at small 't~Hooft coupling. In practice it also provides physical input and a rich testing ground for the weak-coupling limit of the non-perturbative quantum integrable system. Therefore, identifying how the spin-chain constructed in~\cite{OhlssonSax:2011ms, Sax:2012jv, Borsato:2013qpa} emerges on the gauge theory side of the $\AdS_3/\CFT_2$  correspondence is a key outstanding problem.

The $\AdS_3/\CFT_2$ correspondence is different in a number of important ways from its higher-dimensional, higher-supersymmetric cousins. Firstly, unlike $\mathcal{N}=4$ super-Yang-Mills~\cite{Brink:1976bc} or  ABJM theory~\cite{Aharony:2008ug}, the two-dimensional UV gauge theory is \emph{not} conformal: the gauge coupling is dimensionful and the $\CFT_2$ appears as the IR fixed-point of renormalisation group flow. Secondly,  the $\AdS_3/\CFT_2$ dual pair has a large moduli space~\cite{Larsen:1999uk}, while the higher-dimensional duals are parametrised (in the planar limit) only by the value of $\lambda$---the 't~Hooft coupling. Thirdly, the two-dimensional UV gauge theory has, in addition to the (adjoint-valued) gauge 
vector multiplet, a number of (fundamental- and adjoint-valued) hypermultiplets.

The moduli space of the two-dimensional UV gauge theory has two branches: the Coulomb branch and the Higgs branch, which are parametrised by non-zero expectation values of the scalars in vector multiplets and hypermultiplets, respectively. The moduli space metric in the IR is closely related to the UV one because of supersymmetry. The gauge coupling is part of a vector multiplet~\cite{Argyres:1996hc} and so the moduli space metric receives quantum corrections only on the Coulomb branch~\cite{Diaconescu:1997gu}. On the other hand the Fayet-Iliopoulos (FI) parameters~\cite{Fayet:1974jb} and theta angles can be promoted to hypermultiplets and so that they can appear in the Higgs branch moduli space metric~\cite{Diaconescu:1997gu}. Both sets of quantum corrections can be computed exactly because of supersymetry.

Maldacena~\cite{Maldacena:1997re} conjectured that the IR Higgs-branch $\CFT_2$ is dual to string theory on $\AdS_3$. From the string theory point of view this is very natural: the $\AdS_3$ background emerges as the near-horizon limit of a stack of $N_c$ D1- and $N_f$ D5-branes, with the two-dimensional $U(N_c)$ gauge theory living on the D1-branes. In this UV description the Higgs branch parametrises the motion of the D1-branes inside the D5-branes, while on the Coulomb branch the D1-branes separate from the D5-branes. 

An equivalent description of the Higgs branch is provided by the re-combination process of the D-branes as an instanton~\cite{Douglas:1995bn}. The D1-branes are absorbed into the D5-branes forming an instanton (of instanton number $N_c$) in the $SU(N_f)$ gauge theory living on the D5-branes. The low-energy physics~\cite{Douglas:1995bn} matches precisely the ADHM description of instanton moduli space~\cite{Atiyah:1978ri}. Since the moduli-space metric is protected by supersymmetry (it is in fact hyper-K\"ahler), the ADHM sigma-model gives a good description of the Higgs branch $\CFT_2$ for generic instanton configurations. 

However when the size of the instantons shrinks to zero the moduli-space metric becomes singular. Therefore, the $\CFT_2$ states whose wave-function is peaked around such configurations are not well described in terms of a sigma model. There is an alternative description of such states, which can be obtained from the UV gauge-theory Lagrangian by dropping the kinetic terms of the vector multiplet and integrating out the fundamental-valued hypermultiplets, yielding an effective action~\cite{Witten:1997yu,Aharony:1999dw}. This procedure produces new, non-local kinetic terms and interactions for the vector multiplet fields. As a result, such fields no longer have canonical dimensions, rather they have geometric scaling dimensions. The dynamical fields that describe the localised states are adjoint-valued vector multiplets and hypermultiplets, with $1/N_f$ being the effective coupling constant.

In this paper we investigate the $\CFT_2$ states localised near the small-instanton singularity using the description given in~\cite{Witten:1997yu,Aharony:1999dw}.
The gauge-invariant operators in this sub-sector of the $\CFT_2$ can be written in terms of single- and multi-trace operators involving the vector multiplets and adjoint hypermultiplets. We show that, in the large $N_c$ limit, the perturbation series re-arranges itself into a 't~Hooft like expansion in $\lambda\equiv N_c/N_f$ with non-planar corrections suppressed by factors of $1/N_c^2$. As a result, in the planar limit the computation of anomalous dimensions of gauge invariant operators can be reduced to the the spectral problem of a periodic spin-chain with local interactions, much like was found for ${\cal N}=4$ SYM~\cite{Minahan:2002ve}. In a closed $\alg{so}(4)$ sub-sector of the $\CFT_2$ we
calculate the one-loop anomalous dimension of single-trace operators in the planar limit
($N_c \gg 1$) at small 't~Hooft coupling  $N_c/N_f \ll 1$. We show that the resulting
computation reduces in the conventional way~\cite{Minahan:2002ve} to calculating the energy spectrum of
an integrable spin-chain. As far as we are aware, this constitutes the first direct evidence
of integrability on the CFT side of the duality.

In previous works~\cite{OhlssonSax:2011ms, Sax:2012jv, Borsato:2013qpa,Borsato:2014exa}, we have constructed an integrable worldsheet S~matrix which can be used to calculate the perturbative closed string spectrum to all orders in the string tension on the $\AdS_3$ side. Using this construction, the weakly-coupled spin-chain limit of the integrable system was obtained~\cite{OhlssonSax:2011ms,Sax:2012jv}. The representations that appear in this spin-chain are precisely the same as the ones that we find in the spin-chain description of the localised states of the $\CFT_2$ that we find in this paper. Further, in the $\alg{so}(4)$ sub-sector the dilatation
operators of the $\AdS_3$ and $\CFT_2$ agree. These results lead us to suggest that we should identify the perturbative closed string spectrum of the pure R-R $\AdS_3$ theory with the $\CFT_2$ states localised near the small instanton singularity.

%\edited{In this paper we propose to identify the closed-string states whose spectrum is amenable to integrability precisely with the $\CFT_2$ states near the origin of the Higgs branch. In particular, a suitable weak-coupling expansion of the non-local IR action should result in a spin-chain description which is both \emph{local} (at leading order, nearest-neighbour) and \emph{integrable}, like the one constructed in~\cite{OhlssonSax:2011ms, Sax:2012jv, Borsato:2013qpa} by symmetry arguments. The dynamical fields of this IR action will be given by adjoint-valued vector- and hyper-multiplets, with $1/N_f$ being the effective coupling constant of the $\CFT_2$.} In this setting we identify operators with the correct charges and match them to the field content of the spin-chain constructed in~\cite{OhlssonSax:2011ms, Sax:2012jv, Borsato:2013qpa}. In a closed $\algSO(4)$ sub-sector of the $\CFT_2$ we calculate the one-loop anomalous dimension of single-trace operators in the planar limit ($N_c\rightarrow\infty$) at small 't~Hooft coupling $\lambda\equiv N_c/N_f\ll 1$. We show that the resulting computation reduces in the conventional way~\cite{Minahan:2002ve} to calculating the energy spectrum of an integrable spin-chain. As far as we are aware, this constitutes the first direct evidence of integrability on the $\CFT$ side of the duality.

This paper is organised as follows.  In section~\ref{sec:d1d5} we briefly review the UV description of the gauge theory in terms of the D1-D5 system and obtain the UV Lagrangian of the two-dimensional gauge theory by dimensional reduction from six dimensions. In section~\ref{sec:IRcft} we review the description of the $\CFT_2$ states localised near the small-instanton singularity given in~\cite{Witten:1997yu,Aharony:1999dw}. In section~\ref{sec:representations} we show how the field content of that theory fits into representations of $\algPSU(1,1|2)^2$, which is the rigid part of the small $\mathcal{N}=(4,4)$ super-conformal algebra, and how such fields are represented in a spin-chain that matches the one introduced in~\cite{OhlssonSax:2011ms, Sax:2012jv}. In section~\ref{sec:oneloop} we compute the one-loop dilatation operator in the bosonic $\algSO(4)$ subsector of the theory, showing that it corresponds to an integrable Hamiltonian. We conclude in section~\ref{sec:conclusions} and relegate some technical material to the appendices.

\section{The D1-D5 system}
\label{sec:d1d5}

The UV description of the dual pair is encoded in the dynamics of the open strings ending on a stack of D1 and D5-branes,
\begin{center}
  \begin{tabular}{rl}
    $N_f$ D5-branes: & $012345$, \\
    $N_c$ D1-branes: & $01$,
  \end{tabular}
\end{center}
with the directions $2345$ compactified on a four-torus~\cite{Maldacena:1997re}.
The D1- and D5-branes separately preserve supersymmetry corresponding to $\superN=(8,8)$ in two dimensions. However, the intersection of the two stacks of branes only preserve $\superN=(4,4)$. Open strings stretching between the branes give rise to several multiplets of this symmetry. The D1-D1 strings correspond to an $\superN=(8,8)$ $\grpU(N_c)$ vector multiplet, which splits into an $\superN=(4,4)$ vector multiplet and an adjoint hypermultiplet. The D1-D5 strings give bi-fundamental $\grpU(N_c) \times \grpU(N_f)$ hypermultiplets. Finally, the D5-D5 strings give an $\superN=(8,8)$ $\grpU(N_f)$ vector multiplet. In the UV we will take the volume of the 2345 directions to be large. As a result, from the  1+1 dimensional point of view, the D5-D5 strings correspond to non-normalisable modes. Their kinetic terms come with a factor of the 2345 volume and so can be decoupled; the $\grpU(N_f)$ symmetry becomes global. Furthermore, the center-of-mass $\grpU(1)$ vector multiplet also decouples and the gauge group is given by $\grpSU(N_c)$.

The two-dimensional $\superN=(4,4)$ multiplets can be obtained by dimensional reduction of the corresponding six-dimensional $\superN=(1,0)$ multiplets. The six-dimensional supersymmetry algebra has an R-symmetry that we will denote by $\algSU(2)_{\bullet}$.%
\footnote{%
This notation will be useful to make contact with the isometries of~$\AdS_3\times\Sphere^3\times\Torus^4$ as described in~ refs.\cite{Borsato:2014exa,Borsato:2014hja}.
} Upon reduction to two dimension we get an additional $\algSU(2)_{\sL} \oplus\algSU(2)_{\sR}$ symmetry corresponding to rotations in the four compact dimensions. We will also introduce an additional $\algSU(2)_{\circ}$ symmetry under which only the fields in the adjoint hypermultiplet transform. In the brane-system, the $\algSU(2)_{\sL} \oplus \algSU(2)_{\sR}$ symmetry corresponds to rotations in the directions $6$--$9$, while $\algSU(2)_{\bullet} \oplus \algSU(2)_{\circ}$ gives rotations in the directions $2$--$5$.

In the IR the theory has a small $\superN=(4,4)$ superconformal symmetry. The Lie-superalgebra part of this symmetry algebra is given by $\algPSU(1,1|2)^2$. Here $\algSU(2)_{\sL} \oplus \algSU(2)_{\sR}$ plays the role of the R-symmetry, \ie, the two $\algSU(2)$ algebras inside $\algPSU(1,1|2)^2$~\cite{Witten:1997yu}. The algebra $\algSU(2)_{\bullet}$ acts on $\algPSU(1,1|2)^2$ as an outer automorphism, while $\algSU(2)_{\circ}$ commutes with the other symmetries.%
\footnote{%
Our index conventions are as follows. We indicate $\algSO(1,1)$ chiralities by L and R. For the IR R-symmetry $\algSO(4)$ algebra we hence write $\algSO(4)=\algSU(2)_{\sL}\oplus\algSU(2)_{\sR}$, with $\algSU(2)_{\sL,\sR}\subset\algPSU(1,1|2)_{\sL,\sR}$ and we represent the $\algSU(2)$ subalgebras using Greek indices $\dot{\alpha}, \dot{\beta},\dots$ and $\alpha, \beta,\dots$, respectively. The other $\algSO(4)$ is decomposed as $\algSU(2)_{\bullet}\oplus\algSU(2)_{\circ}$ following~\cite{Borsato:2014exa}, with each subalgebra by indices $\dot{a}, \dot{b},\dots$ and $a, b,\dots$, respectively.}

\subsection{Field content in the UV}
The vector multiplet consists of the gauge field $A_{\mu}$, two left-moving fermions $\psi_{\sL}^{\alpha\dot{a}}$, two right-moving fermions $\psi_{\sR}^{\dot{\alpha}\dot{a}}$, four scalars $\phi^{\alpha\dot{\alpha}}$ and three auxiliary fields $D^{\dot{a}\dot{b}}$. Under the global symmetries the fields transform as
\begin{center}
  \begin{tabular}{ccccc}
    \toprule
    & $\algSU(2)_{\sL}$ & $\algSU(2)_{\sR}$ & $\algSU(2)_{\bullet}$ & $\algSU(2)_{\circ}$ \\
    \midrule
    $A_{\mu}$ & $\rep{1}$ & $\rep{1}$ & $\rep{1}$ & $\rep{1}$ \\
    $D^{\dot{a}\dot{b}}$ & $\rep{1}$ & $\rep{1}$ & $\rep{3}$ & $\rep{1}$ \\
    $\phi^{\alpha\dot{\alpha}}$ & $\rep{2}$ & $\rep{2}$ & $\rep{1}$ & $\rep{1}$ \\
    $\psi_{\sL}^{\alpha\dot{a}}$ & $\rep{1}$ & $\rep{2}$ & $\rep{2}$ & $\rep{1}$ \\
    $\psi_{\sR}^{\dot{\alpha}\dot{a}}$ & $\rep{2}$ & $\rep{1}$ & $\rep{2}$ & $\rep{1}$ \\
    \bottomrule
  \end{tabular}
\end{center}
The fundamental hypermultiplet contains two complex scalars $H^{\dot{a}}$, a doublet of left-moving fermions $\lambda_{\sL}^{\dot{\alpha}}$, a doublet of right-moving fermions $\lambda_{\sR}^{\alpha}$, as well as two auxiliary complex scalars $F^{\dot{a}}$. The charges of these fields are given by
\begin{center}
  \begin{tabular}{ccccc}
    \toprule
    & $\algSU(2)_{\sL}$ & $\algSU(2)_{\sR}$ & $\algSU(2)_{\bullet}$ & $\algSU(2)_{\circ}$ \\
    \midrule
    $H^{\dot{a}}$ & $\rep{1}$ & $\rep{1}$ & $\rep{2}$ & $\rep{1}$ \\
    $F^{\dot{a}}$ & $\rep{1}$ & $\rep{1}$ & $\rep{2}$ & $\rep{1}$ \\
    $\lambda_{\sL}^{\dot{\alpha}}$ & $\rep{2}$ & $\rep{1}$ & $\rep{1}$ & $\rep{1}$ \\
    $\lambda_{\sR}^{\alpha}$ & $\rep{1}$ & $\rep{2}$ & $\rep{1}$ & $\rep{1}$ \\
    \bottomrule
  \end{tabular}
\end{center}
The field content of the adjoint hypermultiplet is essentially the same as above , but can be written in terms of real fields by having the full multiplet transforming as a doublet under the global symmetry $\algSU(2)_{\circ}$. We denote the scalars by $T^{{a\dot{a}}}$, the fermions by $\chi_{\sL}^{\dot{\alpha}a}$ and $\chi_{\sR}^{{\alpha}a}$ and the auxiliary field by $G^{{a\dot{a}}}$. The corresponding charges are given by
\begin{center}
  \begin{tabular}{ccccc}
    \toprule
    & $\algSU(2)_{\sL}$ & $\algSU(2)_{\sR}$ & $\algSU(2)_{\bullet}$ & $\algSU(2)_{\circ}$ \\
    \midrule
    $T^{{a\dot{a}}}$ & $\rep{1}$ & $\rep{1}$ & $\rep{2}$ & $\rep{2}$ \\
    $G^{{a\dot{a}}}$ & $\rep{1}$ & $\rep{1}$ & $\rep{2}$ & $\rep{2}$ \\
    $\chi_{\sL}^{\dot{\alpha}a}$ & $\rep{2}$ & $\rep{1}$ & $\rep{1}$ & $\rep{2}$ \\
    $\chi_{\sR}^{{\alpha}a}$ & $\rep{1}$ & $\rep{2}$ & $\rep{1}$ & $\rep{2}$ \\
    \bottomrule
  \end{tabular}
\end{center}

\subsection{Two-dimensional action by dimensional reduction}
To obtain the Lagrangian of the two-dimensional UV field theory one can start with six-dimensional supersymmetric Yang-Mills theory and dimensionally reduce to two dimensions~\cite{Brink:1976bc}. In six dimensions, the fermions in the 
vector multiplet satisfy the symplectic Majorana condition
\begin{equation}
  ( \psi^{\dot{a}} )^* = B \epsilon_{\dot{a}\dot{b}} \psi^{\dot{b}} ,
\end{equation}
and are chiral
\begin{equation}
  \Gamma^{012345} \psi^{\dot{a}} = + \psi^{\dot{a}} .
\end{equation}
The Lagrangian for the vector multiplet is
\begin{equation}
\label{eq:vect-mplet-lag}
  \mathcal{L}_V^{\text{\tiny 6D}} = - \frac{1}{4} \tr F_{\mu\nu} F^{\mu\nu} + \frac{i}{2} \tr \bar{\psi}_{\dot{a}} \slashed{\nabla} \psi^{\dot{a}} + \frac{1}{2} \tr D_{\dot{a}\dot{b}} D^{\dot{a}\dot{b}}.
\end{equation}
The hypermultiplet fermion $\lambda$ is complex and anti-chiral
\begin{equation}
  \Gamma^{012345} \lambda = - \lambda \,.
\end{equation}
%The Lagrangian for a free hypermultiplet is
%\begin{equation}
%  \mathcal{L}_H = -\frac{1}{2} \partial_{\mu} H^{\dag}_{\dot{a}} \partial^{\mu} 
%H^{\dot{a}} + i \bar{\lambda} \slashed{\partial} \lambda + \frac{1}{2} 
%F^{\dag}_{\dot{a}} F^{\dot{a}} .
%\end{equation}
%This is invariant (up to a total derivative) under the supersymmetry transformation
%\begin{equation}
%  \begin{aligned}
 %   \delta H^{\dot{a}} &= + 2 \bar{\epsilon}^{\underline{a}} \lambda , \qquad &
%    \delta \lambda &= - i \epsilon_{\underline{a}} F^{\dot{a}} + i \gamma^{\mu} 
%\epsilon_{\underline{a}} \partial_{\mu} H^{\dot{a}} , \qquad &
%    \delta F^{\dot{a}} &= 2 \bar{\epsilon}^{\underline{a}} \slashed{\partial} 
%\lambda ,
%    \\
%    \delta H^{\dag}_{\dot{a}} &= - 2 \bar{\lambda} \epsilon_{\underline{a}} , \qquad 
%&
%    \delta \bar{\lambda} &= + i F^{\dag}_{\dot{a}} \bar{\epsilon}^{\underline{a}} - i 
%\partial_{\mu} H^{\dag}_{\dot{a}} \bar{\epsilon}_{\underline{a}} \gamma^{\mu}  , 
%\qquad &
%    \delta F^{\dag}_{\dot{a}} &= 2 \partial_\mu \bar{\lambda} \gamma^{\mu} 
%\epsilon_{\underline{a}} .
%  \end{aligned}
%\end{equation}
%The supersymmetry parameter $\epsilon_{\underline{a}}$ is an \emph{anti-chiral} 
%spinor transforming as a doublet under $\algSU(2)_{\sR}$.
The Lagrangian for the hypermultiplet and its couplings to the vector multiplet is
\begin{equation}
  \mathcal{L}_{H}^{\text{\tiny 6D}} = -\frac{1}{2} \nabla_{\mu} H^{\dag}_{\dot{a}} \nabla^{\mu} H^{\dot{a}}
  + i \bar{\lambda} \slashed{\nabla} \lambda
  + i H^{\dag}_{\dot{a}} \bar{\psi}^{\dot{a}} \lambda
  - i \bar{\lambda} \psi^{\dot{a}} H_{\dot{a}}
  + \frac{1}{2} F^{\dag}_{\dot{a}} F^{\dot{a}} 
  + \frac{1}{2} H^{\dag}_{\dot{a}} D^{\dot{a}\dot{b}} H_{\dot{a}} .
\end{equation}
Above we have written the Lagrangian for a hypermultiplet transforming in the fundamental representation of the gauge group. We will write the Lagrangian for the adjoint-valued hypermultiplets in its two-dimensional form below.
%is well-known in the literature (see for example~\cite{Sohnius:1985,Wess:1992cp}) and we will not write it explicitly here. 

%\paragraph{Dimensional reduction to two dimensions.}

When we dimensionally reduce from six to two dimensions the $\algSO(1,5)$ Lorentz symmetry is broken to $\algSO(1,1) \oplus \algSO(4)$ which we write as $\algSO(1,1) \oplus \algSU(2)_{\sL} \oplus \algSU(2)_{\sR}$. The gauge field then splits into a two-dimensional gauge field $A_{\mu}$ and four real scalars $\phi^{\alpha\dot{\alpha}}$ that form a vector of $\algSO(4)$.

Dimensionally reducing the kinetic term for the anti-chiral fermion $\lambda$ to two dimensions we find
\begin{equation}
  i \bar{\lambda} \slashed{\partial} \lambda 
  =
  i \lambda^{\dag} C \slashed{\partial} \lambda
  \to
  i \lambda^{\dag} \partial_t \lambda + i \lambda^{\dag} \hat{\gamma}^{2345} \partial_x \lambda .
\end{equation}
%In the last step we have used that $\lambda$ is 
%anti-chiral (in the six-dimensional sense) and therefore carries index $\alpha_0 = 1$. 
Our gamma-matrix conventions can be found in appendix~\ref{sec:gamma-matrices}.
From this expression we see that worldsheet chirality is correlated with $\algSO(4)$ chirality. Hence, we split the fermion $\lambda$ as
\begin{equation}
  \lambda \to \lambda_{\sL}^{\dot{\alpha}} + \lambda_{\sR}^{\alpha}
\end{equation}
so that the kinetic term becomes
\begin{equation}
  i \lambda_{{\sL}\,\dot{\alpha}}^{\dag} ( \partial_t + \partial_x ) \lambda_{\sL}^{\dot{\alpha}} + i \lambda_{{\sR}\,{\alpha}}^{\dag} ( \partial_t - \partial_x ) \lambda_{\sR}^{{\alpha}} .
\end{equation}
The fermion $\psi^{\dot{a}}$ in the vector multiplet is chiral. The same sort of calculation as above therefore gives
\begin{equation}
  i \bar{\psi}_{\dot{a}} \slashed{\partial} \psi^{\dot{a}}
  \to 
  i \psi_{{\sL}\,\alpha\dot{a}}^{\dag} ( \partial_t + \partial_x ) \psi_{\sL}^{\alpha\dot{a}} + i \psi_{{\sR}\,\dot{\alpha}\dot{a}}^{\dag} ( \partial_t - \partial_x ) \psi_{\sR}^{\dot{\alpha}\dot{a}} .
\end{equation}
The full Lagrangian for the two-dimensional vector multiplet is then given by
\begin{equation}
  \begin{aligned}
    \mathcal{L}_V =
    \tr \Bigl(&
    -\tfrac{1}{4} F_{\mu\nu} F^{\mu\nu} - \tfrac{1}{2} \partial_{\mu} \phi^{\alpha\dot{\alpha}} \partial^{\mu} \phi_{\alpha\dot{\alpha}}
    + i \psi^{\dag}_{\sL\,\alpha\dot{a}} \nabla_+ \psi_{\sL}^{\alpha\dot{a}} + i \psi^{\dag}_{\sR\,\dot{\alpha}\dot{a}} \nabla_- \psi_{\sR}^{\dot{\alpha}\dot{a}}
    \\ &
    - \tfrac{1}{4} \comm{\phi_{\alpha\dot{\alpha}}}{\phi_{\beta\dot{\beta}}} \comm{\phi^{\alpha\dot{\alpha}}}{\phi^{\beta\dot{\beta}}}
    + i \psi^{\dag}_{\sL\,\alpha\dot{a}} \comm{\phi^{\alpha\dot{\alpha}}}{\psi_{\sR\,\dot{\alpha}}{}^{\dot{a}}}
    + i \psi^{\dag}_{\sR\,\dot{\alpha}\dot{a}} \comm{\phi^{\alpha\dot{\alpha}}}{\psi_{\sL\,\alpha}{}^{\dot{a}}}
    + \tfrac{1}{2} D_{\dot{a}\dot{b}} D^{\dot{a}\dot{b}}
    \Bigr) .
  \end{aligned}
\end{equation}
The Lagrangian for the fundamental hypermultiplet and its couplings to the vector multiplet then takes the form
\begin{equation}
\label{eq:fund-hyp-action}
  \begin{aligned}
    \mathcal{L}_H = &
    -\tfrac{1}{2} \nabla_{\mu} H^{\dag}_{\dot{a}} \nabla^{\mu} H^{\dot{a}}
    + i \lambda_{{\sL}\,\dot{\alpha}}^{\dag} \nabla_+ \lambda_{\sL}^{\dot{\alpha}}
    + i \lambda_{{\sR}\,\alpha}^{\dag} \nabla_- \lambda_{\sR}^{\alpha}
    + \tfrac{1}{2} F^{\dag}_{\dot{a}} F^{\dot{a}} 
    \\ &
    - \tfrac{1}{2} H^{\dag}_{\dot{a}} \phi_{\alpha\dot{\alpha}} \phi^{\alpha\dot{\alpha}} H^{\dot{a}}
    + i \lambda_{{\sL}\,\dot{\alpha}}^{\dag} \phi^{\alpha\dot{\alpha}} \lambda_{{\sR}\,\alpha}
    + i \lambda_{{\sR}\,{\alpha}}^{\dag} \phi^{\alpha\dot{\alpha}} \lambda_{{\sL},\dot{\alpha}}
    + \tfrac{1}{2} H^{\dag}_{\dot{a}} D^{\dot{a}\dot{b}} H_{\dot{b}}
    \\ &
    + i H^{\dag}_{\dot{a}} \psi_{\sL}^{\dag\,\alpha \dot{a}} \lambda_{{\sR}\,\alpha}
    + i H^{\dag}_{\dot{a}} \psi_{\sR}^{\dag\,\dot{\alpha}\dot{a}} \lambda_{{\sL}\,\dot{\alpha}}
    - i \lambda_{{\sL}\,\dot{\alpha}}^{\dag} \psi_{\sR}^{\dot{\alpha}\dot{a}} H_{\dot{a}}
    - i \lambda_{{\sR}\,\alpha}^{\dag} \psi_{\sL}^{\alpha\dot{a}} H_{\dot{a}} ,
  \end{aligned}
\end{equation}
and the Lagrangian for the adjoint hypermultiplet the form
\begin{equation}
  \begin{aligned}
    \mathcal{L}_T =
    \tr \Bigl(&
    -\tfrac{1}{2} \nabla_{\mu} T_{a\dot{a}} \nabla^{\mu} T^{a\dot{a}}
    + i \chi^{\dag}_{\sL\,\dot{\alpha}a} \nabla_+ \chi_{\sL}^{\dot{\alpha}a}
    + i \chi^{\dag}_{\sR\,\alpha a} \nabla_- \chi_{\sR}^{\alpha a}
    + \tfrac{1}{2} G_{a\dot{a}} G^{a\dot{a}}
    \\ &
    - \tfrac{1}{2} \comm{\phi_{\alpha\dot{\alpha}}}{T_{a\dot{a}}} \comm{\phi^{\alpha\dot{\alpha}}}{T^{a\dot{a}}}
    + i \chi^{\dag}_{\sL\,\dot{\alpha}a} \comm{\phi^{\alpha\dot{\alpha}}}{\chi_{\sL\alpha}{}^{a}}
    + i \chi^{\dag}_{\sR\,\alpha a} \comm{\phi^{\alpha\dot{\alpha}}}{\chi_{\sR\dot{\alpha}}{}^{a}}
    + \tfrac{1}{2} T_{a\dot{a}} \comm{D^{\dot{a}\dot{b}}}{T^a{}_{\dot{b}}}
    \\ &
    + i T_{a\dot{a}} \comm{\psi_{\sL}^{\dag\,\alpha \dot{a}}}{\chi_{{\sR}\,\alpha}{}^a}
    + i T_{a\dot{a}} \comm{\psi_{\sR}^{\dag\,\dot{\alpha}\dot{a}}}{\chi_{{\sL}\,\dot{\alpha}}^a}
    - i \chi^{\dag}_{{\sL}\,\dot{\alpha}a} \comm{\psi_{\sR}^{\dot{\alpha}\dot{a}}}{T_{\dot{a}}^a}
    - i \chi^{\dag}_{{\sR}\,\alpha a} \comm{\psi_{\sL}^{\alpha\dot{a}}}{T_{\dot{a}}^a}
    \Bigr) .
  \end{aligned}
\end{equation}
We can now write the full two-dimensional UV Lagrangian as
\begin{equation}
  \mathcal{L}_{\text{UV}}(\phi,H,T) = \frac{1}{g_{\text{YM}}^2} \mathcal{L}_V(\phi) + \mathcal{L}_H(H,\phi) + \mathcal{L}_T(T,\phi) ,
\end{equation}
In the above equation we write $\phi$, $H$ and $T$ as short-hand for all fields in these super-multiplets, and the dependence of $\mathcal{L}_{\text{UV}}$ on $g_{\text{YM}}$ has been explicitly written out.

\section{Effective action at the origin of the Higgs branch}
\label{sec:IRcft}

In the previous section we wrote down the two-dimensional $\superN=(4,4)$ SYM Lagrangian $\mathcal{L}_{\text{UV}}$ for the D1-D5 system in the UV. Since $g_{\text{YM}}$ is dimensionful, this theory flows. As explained in~\cite{Witten:1997yu}, the IR is described by two decoupled $\CFT$s: one corresponding to the Coulomb branch and the other to the Higgs branch. Within the context of $\AdS/\CFT$ we will be interested in the Higgs branch $(4,4)$ $\CFT_2$~\cite{Maldacena:1997re}. Witten~\cite{Witten:1997yu} made a proposal for how to describe this $\CFT_2$. The argument is further discussed in~\cite{Aharony:1999dw} and we review it presently. Since $g_{\text{YM}}$ is dimensionful, $\mathcal{L}_V$ 
%~\eqref{eq:vect-mplet-lag} 
is irrelevant and can be dropped while flowing to the IR. The IR Lagrangian is then
\begin{equation}
\mathcal{L}_{\text{IR}}(\phi,H,T) = \mathcal{L}_H(H,\phi) + \mathcal{L}_T(T,\phi) .
\end{equation}
$\mathcal{L}_{\text{IR}}$ is conformal provided we assign the fields in the vector multiplet the following geometric scaling dimensions~\cite{Witten:1997yu}
\begin{equation}
\label{eq:geom-scal}
 \operatorname{dim}(A) = 1 , \qquad
 \operatorname{dim}(\phi) = 1 , \qquad
 \operatorname{dim}(\psi) = 3/2 , \qquad
 \operatorname{dim}(D) = 2 .
\end{equation}
The fields in the hypermultiplets retain their canonical dimensions.

Conventionally, the Higgs branch $\CFT$ can be described as a supersymmetric sigma model on the Higgs branch moduli space. Such a description can be obtained by integrating out the non-dynamical vector multiplet fields in $\mathcal{L}_{\text{IR}}$~\cite{Aharony:1999dw}. However, near the origin of the Higgs branch, this moduli space has a singularity. The degrees of freedom describing states localised near the origin of the Higgs branch are the adjoint-valued vector multiplet and hypermultiplet 
fields~\cite{Witten:1997yu,Aharony:1999dw}. The physics of these localised degrees of freedom is obtained by integrating out from $\mathcal{L}_{\text{IR}}$ the fundamental hypermultiplets $H$
\begin{equation}
  \label{eq:path-integral}
  \int \mathcal{D}\phi \, \mathcal{D} T \, \mathcal{D} H \, e^{i\int d^2x \,  \mathcal{L}_{\text{IR}}}
  =
  \int \mathcal{D}\phi \, \mathcal{D} T \, e^{i\int d^2x ( N_f \mathcal{L}_{\text{eff}}(\phi) + \mathcal{L}_{T}(T,\phi) )}\,.
\end{equation}
Since the adjoint and fundamental hypermultiplets do not couple to one another in $\mathcal{L}_{\text{IR}}$, $\mathcal{L}_{T}$ remains unchanged. On the other hand for the vector multiplet we see that 
this procedure results in what Witten called an "induced gauge theory": the vector multiplet acquires kinetic terms coming exclusively from performing the path integral over the fundamental hypermultiplets. In addition to these kinetic terms, the vector multiplet fields also have higher-order (in the number of fields) interactions determined by the integrating-out procedure. We denote the resulting vector multiplet action by $\mathcal{L}_{\text{eff}}(\phi)$; the Lagrangian governing the adjoint-valued fields is then
\begin{equation}
\label{eq:higgs-origin-lag}
 N_f \mathcal{L}_{\text{eff}}(\phi) + \mathcal{L}_{T}(T,\phi) \,.
\end{equation}
Notice that $\mathcal{L}_{\text{eff}}(\phi)$ appears with an overall factor of $N_f$ which comes from the sum over flavours in the fundamental hypermultiplet loop.\footnote{This action is reminiscent of the Coulomb branch action. The relationship between the two is discussed in refs.~\cite{Seiberg:1999xz,Aharony:1999dw}.} 
In the remainder of this section we discuss in more detail $\mathcal{L}_{\text{eff}}(\phi)$ and how it may be used to perform calculations in the localised sector of the $CFT_2$.

As we mentioned above, integrating out the fundamentals results in the vector multiplet fields becoming dynamical: $\mathcal{L}_{\text{eff}}$ contains non-standard kinetic terms that account for the scaling dimensions given in equation~\eqref{eq:geom-scal}. Conformal invariance then dictates the form of the two-point functions of the fields in the vector multiplet. For example, the dimension one scalar $\phi$ has a two-point function of the form
\begin{equation}
\label{eq:phi-two-pt-fn}
  \braket{\phi^{\alpha\dot{\alpha}}(x) \phi^{\beta\dot{\beta}}(y)} = \frac{C \epsilon^{\alpha\beta} \epsilon^{\dot{\alpha}\dot{\beta}}}{|x-y|^2},
\end{equation}
where $C$ is a normalisation constant. 

The non-standard kinetic terms in $\mathcal{L}_{\text{eff}}$ modify the dynamical degrees of freedom in the vector multiplet. In a theory with a conventional kinetic term, the two-dimensional gauge field $A_{\mu}$ carries no physical degrees of freedom. In  $\mathcal{L}_{\text{eff}}$, on the other hand, $A_{\mu}$ has a non-standard kinetic term, and so the gauge field will now carry one degree of freedom. Similarly, the field $D^{\underline{ab}}$ is auxiliary in the standard Lagrangian, but is a physical field of dimension $2$ in the effective theory. $\superN=(4,4)$ supersymmetry~\cite{Gates:1984nk}, should be powerful enough to determine the kinetic terms of the full vector multiplet from the kinetic term of the scalars $\phi$.\footnote{We are grateful to Chris Hull for discussions about this.}

Witten's description of the states localised near the singularity of the Higgs branch in terms of $\mathcal{L}_{\text{eff}}$ simplifies when $N_c=1$. In that case, it was argued in~\cite{Aharony:1999dw} that the description reduces to a level 
$N_f$ supersymmetric $SU(2)$ WZW model coupled to a charged scalar of charge $Q=(N_f-1)\sqrt{2/N_f}$. In~\cite{Aharony:1999dw}, it is argued that the fields $\phi^{a\dot{a}}$ are composite operators
\begin{equation}
\phi^{\alpha\dot{\alpha}} = e^{\sqrt{2/N_f}\varphi}u^{\alpha\dot{\alpha}}\,,
\end{equation}
where $u$ is the fundamental field of the WZW model and $\varphi$ is the charged scalar. Using the well-known OPEs of $u$ and $\varphi$ one finds
\begin{equation}
\phi^{\alpha\dot{\alpha}}(x)
\phi^{\beta\dot{\beta}}(y)\sim  
\frac{\epsilon^{\alpha\beta}\epsilon^{\dot{\alpha}\dot{\beta}}}{|x-y|^{2-\frac{5}{2N_f}}}\,.
\end{equation}
This is consistent with our expression~\eqref{eq:phi-two-pt-fn} in the large $N_f$ limit.

%%%%%%%%%%%%%%%%%%%%

The effective Lagrangian $\mathcal{L}_{\text{eff}}$ also contains higher-order terms in the vector multiplet fields.
We now describe how these interactions arise from $\mathcal{L}_{\text{IR}}$ and the integrating-out procedure.
In the Lagrangian $\mathcal{L}_{\text{IR}}$ there are interaction vertices involving fields from the fundamental hypermultiplet and fields from the vector multiplet. Hence, the fundamental hypermultiplet mediates interactions between fields of the vector multiplet.
Let us look at an example which will be important in section~\ref{sec:oneloop}. The vector multiplet scalar field~$\phi$ appears quadratically in the Lagrangian $\mathcal{L}_{\text{IR}}$. It interacts with the fundamental hypermultiplet through the terms $H^\dagger \phi\phi H$ and $\lambda^\dagger \phi\lambda$, see equation~\eqref{eq:fund-hyp-action}. As a result, when performing the path integral over the fields $H$ and $\lambda$ in equation~\eqref{eq:path-integral} the resulting functional determinant will yield a non-local quartic effective vertex for $\phi$. This is most transparent in terms of Feynman diagrams:%
\footnote{The effective vertex has a particularly simple expression in terms of Feynman diagrams because the Lagrangian~$\mathcal{L}_{\text{IR}}$ is quadratic in the fields $\lambda$ and $H$ that are integrated out. Hence, the functional integration in equation~\eqref{eq:path-integral} reduces to an one-loop determinant.}
\begin{equation}
\label{eq:effvertex-sec3}
  \begin{tikzpicture}[baseline=-0.5ex]

    \coordinate (i1) at (-0.5cm,-0.5cm);
    \coordinate (i2) at (+0.5cm,-0.5cm);
    \coordinate (o1) at (-0.5cm,+0.5cm);
    \coordinate (o2) at (+0.5cm,+0.5cm);
    \coordinate (v) at (0,0);

    \draw [scalar] (i1) -- (v) -- (o1);
    \draw [scalar] (i2) -- (v) -- (o2);

    \fill [] (v) circle [radius=0.125cm];

  \end{tikzpicture}
  =
  \begin{tikzpicture}[baseline=-0.5ex]

    \coordinate (i1) at (-0.5cm,-0.5cm);
    \coordinate (i2) at (+0.5cm,-0.5cm);
    \coordinate (o1) at (-0.5cm,+0.5cm);
    \coordinate (o2) at (+0.5cm,+0.5cm);

    \coordinate (v1) at (-0.33cm,0);
    \coordinate (v2) at (+0.33cm,0);

    \draw [scalar] (i1) -- (v1) -- (o1);
    \draw [scalar] (i2) -- (v2) -- (o2);

    \draw [hyp scalar,out=71.6,in=108.4] (v1) to (v2);
    \draw [hyp scalar,out=288.4,in=251.6] (v1) to (v2);

    \fill [] (v1) circle [radius=0.0625cm];
    \fill [] (v2) circle [radius=0.0625cm];
  \end{tikzpicture}
+
  \begin{tikzpicture}[baseline=-0.5ex]

    \coordinate (i1) at (-0.5cm,-0.5cm);
    \coordinate (i2) at (+0.5cm,-0.5cm);
    \coordinate (o1) at (-0.5cm,+0.5cm);
    \coordinate (o2) at (+0.5cm,+0.5cm);

    \coordinate (v1) at (-0.25cm,-0.25cm);
    \coordinate (v2) at (+0.25cm,-0.25cm);
    \coordinate (v3) at (+0.25cm,+0.25cm);
    \coordinate (v4) at (-0.25cm,+0.25cm);

    \draw [scalar] (i1) -- (v1);
    \draw [scalar] (i2) -- (v2);
    \draw [scalar] (o2) -- (v3);
    \draw [scalar] (o1) -- (v4);

    \draw [hyp fermion] (v1) -- (v2) -- (v3) -- (v4) -- cycle;

    \fill [] (v1) circle [radius=0.0625cm];
    \fill [] (v2) circle [radius=0.0625cm];
    \fill [] (v3) circle [radius=0.0625cm];
    \fill [] (v4) circle [radius=0.0625cm];
  \end{tikzpicture}\,.
\end{equation}
Here the double lines correspond to $\phi$, the single solid lines to $H$ and the dashed lines to $\lambda$. The fat dot stands for the effective vertex, while the diagrams on the right hand side can be written using the Feynman rules of $\mathcal{L}_{\text{IR}}$.
Equation~\eqref{eq:effvertex-sec3} provides us with an efficient way to compute diagrams involving effective vertices. This information, together with the propagator of equation~\eqref{eq:phi-two-pt-fn} is all we will need to study the one loop mixing of operators in section~\ref{sec:oneloop}. This kind of diagrammatics, as well as the scaling in $N_f$ of equation~\eqref{eq:higgs-origin-lag}, is reminiscent of large-$N$ models~\cite{Coleman:1980nk}.

%%%%%%%%%%%%%%%%

We conclude this section by noting that there are two parameters in the theory: the number of flavours $N_f$ and the number of colours $N_c$. In the effective Lagrangian, $N_f$ plays the role of the Yang-Mills coupling $1/g_{\text{YM}}^2$. Hence, it is natural to consider the weakly coupled 't~Hooft limit, where $N_f, N_c \to \infty$ with $N_c/N_f \ll 1$ fixed.

\section{Superconformal representations of the fields}
\label{sec:representations}
In this section we show how the field content of the theory described above fits in representations of  $\algPSU(1,1|2)^2$, which is the rigid part of the small $\superN=(4,4)$ superconformal algebra, and how these match with the representations used in the spin-chain constructed in~\cite{OhlssonSax:2011ms,Sax:2012jv}. We start by briefly reviewing the $\algU(1,1|2)$ superalgebra.

\subsection{The \texorpdfstring{$\algU(1,1|2)$}{u(1,1|2)} superalgebra}

The $\algU(1,1|2)$ superalgebra consists of the $\algSU(2)$ R-symmetry generators $\gen{R}^a{}_b$, the translation $\gen{P}$, the supercharges $\gen{Q}_a$ and $\dot{\gen{Q}}^a$, the dilatation $\gen{D}$, the special conformal transformation $\gen{K}$, the conformal supercharges $\gen{S}^a$ and $\dot{\gen{S}}_a$, the central charge $\gen{C}$ and the hypercharge $\gen{B}$. We write the $\algSU(2)$ commutation relations as
\begin{equation}
  \comm{\gen{J}^a{}_b}{\gen{R}^c} = + \delta^c_b \gen{R}^a - \tfrac{1}{2} \delta^a_b \gen{R}^c , \qquad
  \comm{\gen{J}^a{}_b}{\gen{R}_c} = - \delta^a_c \gen{R}^b + \tfrac{1}{2} \delta^a_b \gen{R}_c , 
\end{equation}
where $\gen{R}$ is an arbitrary generator. The $\algSU(1,1)$ algebra takes the form
\begin{equation}
  \comm{\gen{K}}{\gen{P}} = 2\gen{D} , \qquad
  \comm{\gen{D}}{\gen{P}} = + \gen{P} , \qquad
  \comm{\gen{D}}{\gen{K}} = - \gen{K} .
\end{equation}
The action of $\gen{P}$ and $\gen{K}$ on the supercharges is given by
\begin{equation}
  \comm{\gen{K}}{\gen{Q}_a} = + \dot{\gen{S}}_a , \qquad
  \comm{\gen{K}}{\dot{\gen{Q}}^a} = + \gen{S}^a , \qquad
  \comm{\gen{P}}{\gen{S}^a} = - \dot{\gen{Q}}^a , \qquad
  \comm{\gen{P}}{\dot{\gen{S}}_a} = - \gen{Q}^a ,
\end{equation}
and the non-trivial anti-commutators by
\begin{equation}
  \begin{aligned}
    \acomm{\gen{Q}_a}{\dot{\gen{Q}}^b} &= \delta_a^b \gen{P} , \qquad &
    \acomm{\gen{Q}_a}{\gen{S}^b} &= - \gen{J}^b{}_a + \delta^b_a ( \gen{D} - \gen{C} ) , \\
    \acomm{\gen{S}^a}{\dot{\gen{S}}_b} &= \delta^a_b \gen{K} , \qquad &
    \acomm{\dot{\gen{Q}}^a}{\dot{\gen{S}}_b} &= - \gen{J}^a{}_b + \delta^a_b ( \gen{D} + \gen{C} ) .
  \end{aligned}
\end{equation}
The (conformal) supercharges carry dimension ($\gen{D}$) and hypercharge ($\gen{B}$)
\begin{equation}
  \begin{aligned}
    \operatorname{dim}(\gen{Q}) = +\tfrac{1}{2} , \qquad &
    \operatorname{dim}(\dot{\gen{Q}}) = +\tfrac{1}{2} , \qquad &
    \operatorname{dim}(\gen{S}) = -\tfrac{1}{2} , \qquad &
    \operatorname{dim}(\dot{\gen{S}}) = -\tfrac{1}{2} ,
    \\
    \operatorname{hyp}(\gen{Q}) = +\tfrac{1}{2} , \qquad &
    \operatorname{hyp}(\dot{\gen{Q}}) = -\tfrac{1}{2} , \qquad &
    \operatorname{hyp}(\gen{S}) = -\tfrac{1}{2} , \qquad &
    \operatorname{hyp}(\dot{\gen{S}}) = +\tfrac{1}{2} .
  \end{aligned}
\end{equation}
Since the hypercharge $\gen{B}$ never appears on the right hand side of the commutation relations we can drop it to obtain the algebra $\algSU(1,1|2)$. If the central charge $\gen{C}$ vanishes we instead get $\algPU(1,1|2)$. Combining these two conditions we obtain the $\algPSU(1,1|2)$ algebra.

\subsection{\texorpdfstring{$\algPSU(1,1|2)_{\sL} \oplus \algPSU(1,1|2)_{\sR}$}{psu(1,1|2) + psu(1,1|2)} decomposition of the fields}

The maximal finite-dimensional subalgebra of the small $\superN=(4,4)$ superconformal algebra is $\algPSU(1,1|2)_{\sL} \oplus \algPSU(1,1|2)_{\sR}$. To distinguish the generators of the two copies of $\algPSU(1,1|2)$ we use subscripts L and R. We also introduce the $\algU(1)$ generators
\begin{equation}
  \begin{gathered}
    \gen{D} = \gen{D}_{\sL} + \gen{D}_{\sR} , \qquad 
    \gen{S} = \gen{D}_{\sL} - \gen{D}_{\sR} , \qquad
    \gen{J} = \gen{J}_{\sL} + \gen{J}_{\sR} , \\
    \gen{J}_{\bullet} = \gen{B}_{\sL} + \gen{B}_{\sR} , \qquad
    \gen{M} = \gen{S} - \gen{J}_{\sL} + \gen{J}_{\sR} .
  \end{gathered}
\end{equation}
The states appearing in the representation%
\footnote{%
We denote by $(-\tfrac{1}{2};\tfrac{1}{2})$ the short representation of $\algPSU(1,1|2)$ having highest weights $-\tfrac{1}{2}$ and $\tfrac{1}{2}$ for its bosonic sub-algebras $\algSU(1,1)$ and $\algSU(2)$, respectively. The representation $(-\tfrac{1}{2};\tfrac{1}{2})^*$ differs from $(-\tfrac{1}{2};\tfrac{1}{2})$ by having a  fermionic, rather than bosonic, highest weight state.
}
 $(-\tfrac{1}{2};\tfrac{1}{2})_{\sL} \otimes (-\tfrac{1}{2};\tfrac{1}{2})_{\sR}$ of this algebra  were used to construct the all-loop massive S~matrix of the $\AdS_3\times\Sphere^3\times\Torus^4$ superstring~\cite{OhlssonSax:2011ms}. We collect them in table~\ref{tab:psu112-2-states}.
To compare these states with the ones appearing in the field theory, we identify the $\algSU(2)_{\sL} \oplus \algSU(2)_{\sR} \subset \algPSU(1,1|2)_{\sL} \oplus \algPSU(1,1|2)_{\sR}$ with the $\algSO(4)$ we got by reducing from six to two dimensions. We further let $\gen{D}$ measure the dimension and $\gen{S}$ the Lorentz spin. Finally we identify $\gen{J}_{\bullet}$ with the $\algU(1)$ part of the $\algSU(2)_{\bullet}$ R-symmetry. We will now see how the states in the $(-\tfrac{1}{2};\tfrac{1}{2})_{\sL} \otimes (-\tfrac{1}{2};\tfrac{1}{2})_{\sR}$ match the field theory states.
\begin{table}
  \centering
  \begin{tabular}{ccccccc}
    \toprule
    Spin-chain & $D$ & $S$ & $J_{\sL}$ & $J_{\sR}$ & $J_{\bullet}$ & $\CFT_2$ \\
    \midrule
    $\phi_{\sL}^{\pm} \phi_{\sR}^{\pm}$ & $1$ & $0$ & $\pm \tfrac{1}{2}$ & $\pm \tfrac{1}{2}$ & $0$ & $\phi^{\alpha\dot{\alpha}}$\\
    \midrule
    $\phi_{\sL}^{\pm} \psi_{\sR}^{\pm}$ & $\tfrac{3}{2}$ & $-\tfrac{1}{2}$ & $\pm \tfrac{1}{2}$ & $0$ & $\pm \tfrac{1}{2}$ & $\psi_{\sR}^{\dot{\alpha}\dot{a}}$ \\
    $\psi_{\sL}^{\pm} \phi_{\sR}^{\pm}$ & $\tfrac{3}{2}$ & $+\tfrac{1}{2}$ & $0$ & $\pm \tfrac{1}{2}$ & $\pm \tfrac{1}{2}$ & $\psi_{\sL}^{\alpha\dot{a}}$ \\
    \midrule
    $\psi_{\sL}^{\pm} \psi_{\sR}^{\pm}$ & $2$ & $0$ & $0$ & $0$ & $\pm 1$ & \multirow{2}{*}{$\epsilon^{\mu\nu}F_{\mu\nu}$, $D^{\dot{a}\dot{b}}$} \\
    $\psi_{\sL}^{\pm} \psi_{\sR}^{\mp}$ & $2$ & $0$ & $0$ & $0$ & $0$ & \\
    \midrule
    $\mathcal{D}_{\sL} \phantom{\mathcal{D}_{\sR}}$ & $1$ & $+1$ & $0$ & $0$ & $0$ & $\nabla_+$ \\
    $\phantom{\mathcal{D}_{\sL}} \mathcal{D}_{\sR}$ & $1$ & $-1$ & $0$ & $0$ & $0$ & $\nabla_-$\\
    \bottomrule
  \end{tabular}
  \caption{Charges of the states in the representation $(-\tfrac{1}{2};\tfrac{1}{2})_{\sL} \otimes (-\tfrac{1}{2};\tfrac{1}{2})_{\sR}$ and their relation to the degrees of freedom of the spin-chain and $\CFT_2$. We use $\phi^{\pm}_{\sL/\sR}$ to denote the $\algSU(2)_{\sL/\sR}$ doublet of scalars in the $(-\tfrac{1}{2};\tfrac{1}{2})_{\sL/\sR}$ representation, while $\psi^{\pm}_{\sL/\sR}$ denotes the two fermions. The $\algSU(1,1)_{\sL/\sR}$ descendants are obtained by acting on these states with $\mathcal{D}_{\sL/\sR}$.}
  \label{tab:psu112-2-states}
\end{table}

First we have four scalars with dimension $1$ transforming as a bispinor under $\algSU(2)_{\sL} \oplus \algSU(2)_{\sR}$. These correspond to the fields $\phi^{\alpha\dot{\alpha}}$. There are eight dimension $\tfrac{3}{2}$ fermions that have charge $\pm\tfrac{1}{2}$ under $\gen{J}_{\bullet}$. The fermions with spin $+\tfrac{1}{2}$ are doublets under $\algSU(2)_{\sR}$ and those with spin $-\tfrac{1}{2}$ are doublets under $\algSU(2)_{\sL}$. This matches the chiral and anti-chiral spinors $\psi_{\sL}^{\alpha\dot{a}}$ and $\psi_{\sR}^{\dot{\alpha}\dot{a}}$. The $\algSU(1,1)_{\sL}$ and $\algSU(1,1)_{\sR}$ descendants are obtained by acting with $\mathcal{D}_{\sL}$ and $\mathcal{D}_{\sR}$. These correspond to the left- and right-moving covariant derivatives $\nabla_t \pm \nabla_x$.

Finally, there are four bosons of dimension $2$, which are only charged under $\algSU(2)_{\bullet}$, which decompose into a singlet and a triplet. The triplet can be identified with the auxiliary field $D^{\dot{a}\dot{b}}$, while the singlet comes from the field strength $\epsilon^{\mu\nu}F_{\mu\nu} = 2\epsilon^{\mu\nu}\nabla_{\mu} A_{\nu}$. Hence,  the field strength multiplet perfectly fits into $(-\tfrac{1}{2};\tfrac{1}{2})_{\sL} \otimes (-\tfrac{1}{2};\tfrac{1}{2})_{\sR}$.

The multiplet discussed here is very similar to the field strength multiplet of $\superN=4$ SYM (see \eg ref.~\cite{Beisert:2004ry}). However, the two-dimensional field strength has only a single component $F_{tx} = -F_{xt}$. Further, as seen above the field strength multiplet contains the normally auxiliary field $D^{\dot{a}\dot{b}}$ which now is dynamical. In table~\ref{tab:psu112-2-states} we also summarise the identifications of the spin-chain and $\CFT_2$ variables.

The states in the adjoint hypermultiplet can also be organised into $\algPSU(1,1|2)_{\sL} \oplus \algPSU(1,1|2)_{\sR}$ representations. However, in this case the representations are \emph{reducible}. The scalars $T^{\pm\pm}$ have vanishing dimensions, spin and $\algSU(2)_{\sL} \oplus \algSU(2)_{\sR}$ charge. Hence, they form singlets under $\algPSU(1,1|2)_{\sL} \oplus \algPSU(1,1|2)_{\sR}$. However, they have spin $1/2$ under both $\algSU(2)_{\bullet}$ and $\algSU(2)_{\circ}$.

The remaining states of the hypermultiplet are charged under either $\algPSU(1,1|2)_{\sL}$ or $\algPSU(1,1|2)_{\sR}$. The left-moving fermions $\chi_{\sL}^{+\pm}$ are the highest-weight states of $(-\tfrac{1}{2};\tfrac{1}{2})_{\sL}^* \otimes 1_{\sR}^{\pm}$, where $1_{\sR}^{\pm}$ denotes two $\algPSU(1,1|2)_{\sR}$ singlets that form a doublet under $\algSU(2)_{\circ}$. The other states in these representations are given by $\chi_{\sL}^{-\pm}$ and $\nabla_+ T^{\pm\pm}$ as well as the $\algSU(1,1)_{\sL}$ descendants obtained by acting with $\nabla_+$. Similarly, the right-moving fermions $\chi_{\sR}^{+\pm}$ are highest-weight states of the modules $1_{\sL}^{\pm} \otimes (-\tfrac{1}{2};\tfrac{1}{2})_{\sR}^*$, which further contain $\chi_{\sR}^{-\pm}$ and $\nabla_- T^{\pm\pm}$ as well as the states obtained by the action of $\nabla_-$. The charges of the states in these multiplets and the identification of spin-chain and $\CFT_2$ degrees of freedom are summarised in table~\ref{tab:psu112-2-states-hyper}. This structure matches the reducible spin chain constructed in ref.~\cite{Sax:2012jv}.
\begin{table}
  \centering
  \begin{tabular}{cccccccc}
    \toprule
    spin-chain & $D$ & $S$ & $J_{\sL}$ & $J_{\sR}$ & $J_{\bullet}$ & $J_{\circ}$ & $\CFT_2$ \\
    \midrule
    $\tilde{\psi}_{\sL}^{\pm} 1_{\sR}^{\pm}$ & $\tfrac{1}{2}$ & $+\tfrac{1}{2}$ & $\pm \tfrac{1}{2}$ & $0$ & $0$ & $\pm \tfrac{1}{2}$ & $\chi_{\sL}^{\dot{\alpha}a}$\\
    $\tilde{\phi}_{\sL}^{\pm} 1_{\sR}^{\pm}$ & $1$ & $+1$ & $0$ & $0$ & $\pm\tfrac{1}{2}$ & $\pm\tfrac{1}{2}$ & $\nabla_+ T^{{a\dot{a}}}$ \\
    \midrule
    $1_{\sL}^{\pm} \tilde{\psi}_{\sR}^{\pm}$ & $\tfrac{1}{2}$ & $-\tfrac{1}{2}$ & $0$ & $\pm \tfrac{1}{2}$ & $0$ & $\pm \tfrac{1}{2}$ & $\chi_{\sR}^{{\alpha}a}$\\
    $1_{\sR}^{\pm} \tilde{\phi}_{\sR}^{\pm}$ & $1$ & $-1$ & $0$ & $0$ & $\pm\tfrac{1}{2}$ & $\pm\tfrac{1}{2}$ & $\nabla_- T^{{a\dot{a}}}$ \\
    \midrule
    $\mathcal{D}_{\sL} \phantom{\mathcal{D}_{\sR}}$ & $1$ & $+1$ & $0$ & $0$ & $0$ & $0$ & $\nabla_+$ \\
    $\phantom{\mathcal{D}_{\sL}} \mathcal{D}_{\sR}$ & $1$ & $-1$ & $0$ & $0$ & $0$ & $0$ & $\nabla_-$\\
    \bottomrule
  \end{tabular}
  \caption{Charges of the states in the representation $1_{\sL}^{\pm} \otimes (-\tfrac{1}{2};\tfrac{1}{2})_{\sR}^*$ and $(-\tfrac{1}{2};\tfrac{1}{2})_{\sL}^* \otimes 1_{\sR}^{\pm}$ and their relation to the degrees of freedom of the spin-chain and $\CFT_2$. The fermions $\tilde{\psi}^{\pm}_{\sL/\sR}$ form an $\algSU(2)_{\sL/\sR}$ doublet and $\tilde{\phi}^{\pm}_{\sL/\sR}$ denote the two bosons in the $(-\tfrac{1}{2};\tfrac{1}{2})_{\sL/\sR}^*$ representations. The $\algSU(1,1)_{\sL/\sR}$ descendants are again obtained by acting with $\mathcal{D}_{\sL/\sR}$.}
  \label{tab:psu112-2-states-hyper}
\end{table}

\subsection{Spin-chain states}

We can now construct local operators using the states in the field strength multiplet and adjoint hypermultiplet. The simplest such state is the 1/2-BPS ground state
\begin{equation}
  \label{eq:vacuum}
  \tr \left[( \phi^{++} )^L\right] .
\end{equation}
Starting with this state we can obtain excited states by replacing the field at a site by any other adjoint field. Let us first restrict ourselves to the fields in the field strength multiplet. Such an operator will consist of $L$ sites each containing a state from the $(-\tfrac{1}{2};\tfrac{1}{2})_{\sL} \otimes (-\tfrac{1}{2};\tfrac{1}{2})_{\sR}$ representation of $\algPSU(1,1|2)_{\sL} \oplus \algPSU(1,1|2)_{\sR}$. This is exactly the operators that appear in the integrable spin-chain constructed in~\cite{OhlssonSax:2011ms}. The Hamiltonian acting on the spin-chain is given by the charge $\gen{H} = \gen{D} - \gen{J}$. In the next section we will calculate the one-loop correction to this Hamiltonian in a closed sub-sector.

The ground state~\eqref{eq:vacuum} is the highest weight state in the short $(-\tfrac{L}{2};\tfrac{L}{2})_{\sL} \otimes (-\tfrac{L}{2};\tfrac{L}{2})_{\sR}$ representation of $\algPSU(1,1|2)_{\sL} \oplus \algPSU(1,1|2)_{\sR}$, and carries $\algU(1)$ charges $D-J = 0$ and $M = 0$. The lowest-lying excitations, \ie, the excitations with lowest eigenvalue for $\gen{D}-\gen{J}$, created by the fields in the field strength multiplet can be grouped into two sets. Replacing one of the fields $\phi^{++}$ in the ground state by any of the fields
\begin{equation}
  \phi^{-+} , \quad
  \nabla_+ \phi^{-+} , \quad
  \psi_{\sL}^{+ +} , \quad
  \psi_{\sL}^{- +} 
\end{equation}
results in a state with $D - J = 1$ and $M = +1$. We can also consider the excitations
\begin{equation}
  \phi^{+-} , \quad
  \nabla_- \phi^{+-} , \quad
  \psi_{\sR}^{+ +} , \quad
  \psi_{\sR}^{- +} 
\end{equation}
which carry charges $D - J = 1$ and $M = -1$. These two sets of excitations form two multiplets of the centrally extended $\algPSU(1|1)^4_{\text{c.e.}}$ used to construct the S matrix in~\cite{Borsato:2013qpa}. The remaining fields in the field strength multiplet can be though of as composite excitations constructed out of the eight fundamental excitations.

Let us now consider excitations created from fields in the adjoint hypermultiplet. The four bosonic and four fermionic fields
\begin{equation}
  T^{\pm\pm} , \quad
  \chi_{\sL}^{+\pm} , \quad
  \chi_{\sR}^{+\pm} ,
\end{equation}
all carry charges $D - J = 0$ and $M=0$. They can be inserted into the operator with no cost of energy, and hence correspond to gapless excitations. Since the scalars $T^{a\dot{a}}$ form singlets under $\algPSU(1,1|2)_{\sL} \oplus \algPSU(1,1|2)_{\sR}$ the resulting spin-chain states are exactly of the type that appears in the reducible spin-chain discussed in~\cite{Sax:2012jv}. The excitations fit perfectly in the massless $\algPSU(1|1)^4_{\text{c.e.}} \oplus \algSU(2)_{\circ}$ representation discussed in~\cite{Borsato:2014exa,Borsato:2014hja}. While the operators containing these fields form reducible short representations at tree-level, we expect them to join into long irreducible representations once we include the interactions. 
A similar mechanism explains the fact that at zero coupling this theory seems to have too many chiral primaries with respect to the supergravity dual or to the symmetric product orbifold CFT~\cite{deBoer:1998ip}. Naively, arbitrary products of $\chi_{\sL}^{+\pm},  \chi_{\sR}^{+\pm}$ give chiral primary composite operators. However, all but few of these are accidental chiral primaries, whose dimension is corrected at non-zero coupling~\cite{Sax:2012jv}.
 We hope return to the investigation of the exact details of these mechanisms and of the chiral ring in the future.

\section{One-loop dilatation operator in the  \texorpdfstring{$\algSO(4)$}{so(4)} sector}
\label{sec:oneloop}

In this section we will compute, in the planar limit, the one-loop anomalous dimensions of operators in the effective $\CFT_2$ proposed by Witten. We restrict ourselves to operators made up of scalar fields from the vector multiplet, which form a one-loop closed subsector charged under $\algSO(4)$.%
\footnote{%
We will return to the analogous computation for the complete theory in an upcoming work.
}
We find that the computation of these anomalous dimensions can be re-phrased in terms of finding the spectrum of a nearest-neighbour Hamiltonian for a homogenous $\algSO(4)$ spin chain, in a manner similar to the seminal work of Minahan and Zarembo~\cite{Minahan:2002ve}. As we will show explicitly, such an Hamiltonian has precisely the right form to be \emph{integrable}. 

Witten's prescription for the dynamics for states localised near the origin of the Higgs branch is closely related to the ``large-N'' approach\footnote{See~\cite{Coleman:1980nk} for reviews and a more extensive list of references.} widely used in the study of sigma models~\cite{D'Adda:1978uc}. In the present setting it is $N_f$ that plays the role of ``large-N'', while taking $N_c$ large will ensure that we can restrict to single-trace operators. Let us then consider single-trace operators made out of scalar fields from the vector multiplet,
\begin{equation}
  \mathcal{O}^A = \tr \bigl(
  \phi^{\alpha_1\dot{\alpha}_1} \phi^{\alpha_2\dot{\alpha}_2} \dotsb \phi^{\alpha_{J}\dot{\alpha}_{J}}
  \bigr) .
\end{equation}
Since these operators are charged only under $\algSO(4) = \algSU(2)_{\sL} \oplus \algSU(2)_{\sR}$ we refer to it as the $\algSO(4)$ sector.
Our goal is to find the one-loop dilatation operator acting on these operators. To extract the dilatation operator we consider the one-loop corrections to the local operators. These corrections are generally UV divergent, which means we need to introduce counter-terms and renormalised operators
\begin{equation}
  \mathcal{O}_{I,\text{ren}} = \mathcal{Z}_I{}^J \mathcal{O}_{J,\text{bare}} , \qquad
  \mathcal{Z} = 1 + \lambda^2 \mathcal{Z}_2 + \dotsb .
\end{equation}
The matrix $\mathcal{Z}$ cancels the UV divergences in the one-loop corrections, and introduces a mixing between different operators. In a perturbative calculation it can be extracted as minus the divergent piece of the sum of all Feynman diagrams. In dimensional regularisation we consider the theory in $D = d + 2\epsilon$ dimensions and divergences appear as poles in $\epsilon$. For non-zero $\epsilon$ the coupling constant is dimensionful. Hence we introduce as mass parameter $\mu$ so that the full coupling constant can be written as $\mu^{2\epsilon} \lambda$. The corrections to the dilatation operator can then be extracted as
\begin{equation}
  \delta\gen{D} = \mu \frac{d}{d \mu} \log \mathcal{Z}(\mu^{2\epsilon} \lambda, \epsilon) \Bigl|_{\epsilon=0} .
\end{equation}
%This expression clearly only makes sense if $\log \mathcal{Z}$ contains at most simple poles. Note that this is simply a condition on the renormalisability of the theory.

In the planar limit, the one-loop dilatation operator acts locally on two neighbouring fields of the operators. From $\algSO(4)$ symmetry we expect it to take the form
\begin{equation}
  \sum_{k=1}^{L} ( \delta_{a_1}^{b_1} \delta_{\dot{a}_1}^{\dot{b}_1} ) \dotsb ( \delta_{a_{k-1}}^{b_{k-1}} \delta_{\dot{a}_{k-1}}^{\dot{b}_{k-1}} )
  M_{a_k \dot{a}_k a_{k+1} \dot{a}_{k+1}}^{b_k \dot{b}_k b_{k+1} \dot{b}_{k+1}}
  ( \delta_{a_{k+2}}^{b_{k+2}} \delta_{\dot{a}_{k+2}}^{\dot{b}_{k+2}} ) \dotsb ( \delta_{a_{J}}^{b_{J}} \delta_{\dot{a}_{J}}^{\dot{b}_{J}} ) ,
\end{equation}
where $M$ is an $\algSO(4)$ invariant tensor, which can be built out of the identity $\mathds{1}$, the permutation $\mathds{P}$ and the trace $\mathds{K}$.
The dilatation operator should preserve supersymmetry. In particular it should annihilate chiral primary operators such as \eqref{eq:vacuum},
since the dimension of these operators is protected. Since this operator is symmetric and traceless, we can impose supersymmetry by requiring the tensor $M$ to take the form
\begin{equation}
  M = c_1 ( \mathds{1} - \mathds{P} ) + c_2 \mathds{K}\,,
\end{equation}
for some constants $c_1$ and $c_2$. Hence, we only need to consider Feynman diagrams that act non-trivially in flavour space.

\subsection{``One-loop'' Feynman diagrams}

We now proceed to compute the Feynman diagrams in order to determine the form of the spin-chain Hamiltonian for the $\algSO(4)$ sector involving $\phi^{\alpha\dot{\alpha}}$.
The dynamics of the field $\phi$ is given by the effective Lagrangian $\mathcal{L}_{\text{eff}}$. This has non-standard propagators and non-local effective vertices. As we have discussed in section~\ref{sec:IRcft}, such propagators are fixed by conformal invariance, \textit{cfr.}\ equation~\eqref{eq:phi-two-pt-fn}, and the effective interaction vertices are given by one-loop diagrams of~$\mathcal{L}_{\text{IR}}$, like in~\eqref{eq:effvertex-sec3}. This provides a systematic expansion of $\mathcal{L}_{\text{eff}}$ in Feynman diagrams.

We are interested in the limit of a large number of colours $N_c$ and flavours $N_f$, with the 't~Hooft coupling $\lambda = N_c / N_f$ fixed and small. The (effective) diagrams that contribute to the two-point function at one loop and are divergent by power counting are
\begin{equation}\label{eq:D-one-loop-diagrams}
  \begin{tikzpicture}[baseline=-0.5ex]

    \coordinate (i1) at (0,-0.75cm);
    \coordinate (i2) at (1cm,-0.75cm);

    \coordinate (o1) at (0,0.75cm);
    \coordinate (o2) at (1cm,0.75cm);

    \coordinate (v) at (0.5cm,0cm);

    \draw [scalar,out=90,in=225] (i1) to (v);
    \draw [scalar,out=90,in=315] (i2) to (v);

    \draw [scalar,out=135,in=270] (v) to (o1);
    \draw [scalar,out=45,in=270] (v) to (o2);

    \fill [] (v) circle [radius=0.125cm];

    \draw [thick,rounded corners=2pt,fill=gray] ($(i1)+(-0.25cm,-0.125cm)$) -- ($(i1)+(-0.25cm,0)$) -- ($(i2)+(0.25cm,0)$) -- ($(i2)+(0.25cm,-0.125cm)$) -- cycle;
  \end{tikzpicture}
\qquad
  \begin{tikzpicture}[baseline=-0.5ex]

    \coordinate (i1) at (0,-0.75cm);
    \coordinate (i2) at (1cm,-0.75cm);

    \coordinate (o1) at (0,0.75cm);
    \coordinate (o2) at (1cm,0.75cm);

    \coordinate (v1) at (0,0);
    \coordinate (v2) at (1cm,0);

    \draw [scalar] (i1) to (v1) to (o1);
    \draw [scalar] (i2) to (v2) to (o2);

    \draw [gluon] (v1) to (v2);

    \fill [] (v1) circle [radius=0.125cm];
    \fill [] (v2) circle [radius=0.125cm];

    \draw [thick,rounded corners=2pt,fill=gray] ($(i1)+(-0.25cm,-0.125cm)$) -- ($(i1)+(-0.25cm,0)$) -- ($(i2)+(0.25cm,0)$) -- ($(i2)+(0.25cm,-0.125cm)$) -- cycle;
  \end{tikzpicture}
\qquad
  \begin{tikzpicture}[baseline=-0.5ex]

    \coordinate (i1) at (0,-0.75cm);
    \coordinate (i2) at (1cm,-0.75cm);

    \coordinate (o1) at (0,0.75cm);
    \coordinate (o2) at (1cm,0.75cm);

    \coordinate (v1) at (0.5cm,-0.3cm);
    \coordinate (v2) at (0.5cm,0.3cm);

    \draw [scalar,out=90,in=180] (i1) to (v1);
    \draw [scalar,out=90,in=0] (i2) to (v1);

    \draw [scalar,out=180,in=270] (v2) to (o1);
    \draw [scalar,out=0,in=270] (v2) to (o2);

    \draw [gluon] (v1) to (v2);

    \fill [] (v1) circle [radius=0.125cm];
    \fill [] (v2) circle [radius=0.125cm];

    \draw [thick,rounded corners=2pt,fill=gray] ($(i1)+(-0.25cm,-0.125cm)$) -- ($(i1)+(-0.25cm,0)$) -- ($(i2)+(0.25cm,0)$) -- ($(i2)+(0.25cm,-0.125cm)$) -- cycle;
  \end{tikzpicture}
\qquad
  \begin{tikzpicture}[baseline=-0.5ex]

    \coordinate (i) at (0,-0.75cm);
    \coordinate (o) at (0,0.75cm);
    \coordinate (v) at (0,0);

    \draw [scalar] (i) to (v) to (o);

    \draw [thick,fill=white] (v) circle [radius=0.25cm];
    \draw [very thick,pattern=north east lines] (v) circle [radius=0.25cm];

    \draw [thick,rounded corners=2pt,fill=gray] ($(i)+(-0.25cm,-0.125cm)$) -- ($(i)+(-0.25cm,0)$) -- ($(i)+(0.25cm,0)$) -- ($(i)+(0.25cm,-0.125cm)$) -- cycle;
  \end{tikzpicture} \, ,
\end{equation}
where the straight and wiggly double lines indicate the effective propagators of the scalars and gluons from the vector multiplet, and the fat dots represent effective vertices obtained by one-loop sub-diagrams with a field from the fundamental hypermultiplet running in the loop. The large blob in the last diagram represents the one-loop self energy. As we saw in eq.~\eqref{eq:effvertex-sec3}, the four-scalar interaction comes from the diagrams
\begin{equation}
\label{eq:effvertex1}
  \begin{tikzpicture}[baseline=-0.5ex]

    \coordinate (i1) at (-0.5cm,-0.5cm);
    \coordinate (i2) at (+0.5cm,-0.5cm);
    \coordinate (o1) at (-0.5cm,+0.5cm);
    \coordinate (o2) at (+0.5cm,+0.5cm);
    \coordinate (v) at (0,0);

    \draw [scalar] (i1) -- (v) -- (o1);
    \draw [scalar] (i2) -- (v) -- (o2);

    \fill [] (v) circle [radius=0.125cm];

  \end{tikzpicture}
  =
  \begin{tikzpicture}[baseline=-0.5ex]

    \coordinate (i1) at (-0.5cm,-0.5cm);
    \coordinate (i2) at (+0.5cm,-0.5cm);
    \coordinate (o1) at (-0.5cm,+0.5cm);
    \coordinate (o2) at (+0.5cm,+0.5cm);

    \coordinate (v1) at (-0.33cm,0);
    \coordinate (v2) at (+0.33cm,0);

    \draw [scalar] (i1) -- (v1) -- (o1);
    \draw [scalar] (i2) -- (v2) -- (o2);

    \draw [hyp scalar,out=71.6,in=108.4] (v1) to (v2);
    \draw [hyp scalar,out=288.4,in=251.6] (v1) to (v2);

    \fill [] (v1) circle [radius=0.0625cm];
    \fill [] (v2) circle [radius=0.0625cm];
  \end{tikzpicture}
+
  \begin{tikzpicture}[baseline=-0.5ex]

    \coordinate (i1) at (-0.5cm,-0.5cm);
    \coordinate (i2) at (+0.5cm,-0.5cm);
    \coordinate (o1) at (-0.5cm,+0.5cm);
    \coordinate (o2) at (+0.5cm,+0.5cm);

    \coordinate (v1) at (-0.25cm,-0.25cm);
    \coordinate (v2) at (+0.25cm,-0.25cm);
    \coordinate (v3) at (+0.25cm,+0.25cm);
    \coordinate (v4) at (-0.25cm,+0.25cm);

    \draw [scalar] (i1) -- (v1);
    \draw [scalar] (i2) -- (v2);
    \draw [scalar] (o2) -- (v3);
    \draw [scalar] (o1) -- (v4);

    \draw [hyp fermion] (v1) -- (v2) -- (v3) -- (v4) -- cycle;

    \fill [] (v1) circle [radius=0.0625cm];
    \fill [] (v2) circle [radius=0.0625cm];
    \fill [] (v3) circle [radius=0.0625cm];
    \fill [] (v4) circle [radius=0.0625cm];
  \end{tikzpicture}\,,
\end{equation}
while the gluon-scalar interaction is given by
\begin{equation}
\label{eq:effvertex2}
  \begin{tikzpicture}[baseline=-0.5ex]

    \coordinate (i1) at (120 : 0.5cm);
    \coordinate (i2) at (240 : 0.5cm);
    \coordinate (o) at (0 : 0.5cm);
    \coordinate (v) at (0,0);

    \draw [scalar] (i1) -- (v);
    \draw [scalar] (i2) -- (v);
    \draw [gluon] (o) -- (v);

    \fill [] (v) circle [radius=0.125cm];
  \end{tikzpicture}
  =
  \begin{tikzpicture}[baseline=-0.5ex]

    \coordinate (i1) at (120 : 0.5cm);
    \coordinate (i2) at (240 : 0.5cm);
    \coordinate (v1) at (0,0);
    \coordinate (v2) at (0.5cm,0);
    \coordinate (o) at (0.75cm,0);

    \draw [scalar] (i1) -- (v1);
    \draw [scalar] (i2) -- (v1);
    \draw [gluon] (o) -- (v2);

    \draw [hyp scalar,out=60,in=120] (v1) to (v2);
    \draw [hyp scalar,out=300,in=240] (v1) to (v2);

    \fill [] (v1) circle [radius=0.0625cm];
    \fill [] (v2) circle [radius=0.0625cm];
  \end{tikzpicture}
  +
  \begin{tikzpicture}[baseline=-0.5ex]

    \coordinate (i1) at (120 : 0.5cm);
    \coordinate (i2) at (240 : 0.5cm);
    \coordinate (v1) at (120 : 0.25cm);
    \coordinate (v2) at (240 : 0.25cm);
    \coordinate (v3) at (0.25cm,0);
    \coordinate (o) at (0.5cm,0);

    \draw [scalar] (i1) -- (v1);
    \draw [scalar] (i2) -- (v2);
    \draw [gluon] (o) -- (v3);

    \draw [hyp fermion] (v1) -- (v2);
    \draw [hyp fermion] (v2) -- (v3);
    \draw [hyp fermion] (v3) -- (v1);

    \fill [] (v1) circle [radius=0.0625cm];
    \fill [] (v2) circle [radius=0.0625cm];
    \fill [] (v3) circle [radius=0.0625cm];

  \end{tikzpicture}\,.
\end{equation}
Above, the single dashed and solid lines correspond to fundamental fermions and scalars, respectively. We refer to the diagrams in~\eqref{eq:D-one-loop-diagrams} as \emph{one-loop diagrams}, because they all give a contribution proportional to the coupling $\lambda = N_c / N_f$. In practice, to compute them we need to plug in the effective vertices (\ref{eq:effvertex1}), (\ref{eq:effvertex2}), yielding several loops for each diagram. Let us see in more detail how this counting of~$\lambda$ goes for the diagrams in \eqref{eq:D-one-loop-diagrams}. We first note that the above vertices all contain a single loop of fields from the fundamental hypermultiplet, which gives an overall factor of $N_f$. Furthermore, the effective action for the fields in the vector multiplet is proportional to $N_f$. Hence, each double line propagator in~\eqref{eq:D-one-loop-diagrams} gives a factor $1/N_f$. Finally, in each diagram there is a single sum over colour indices in the loop, which gives a factor of $N_c$. Since the number of effective vertices in each diagram is one less than the number of internal propagators the overall factor for each diagram is $N_c / N_f$.

\begin{figure}
  \centering
  
  \subfloat[$N_c/N_f$]{
    \begin{tikzpicture}
      \coordinate (i1) at (0.0cm,0.0cm);
      \coordinate (i2) at (1.5cm,0.0cm);

      \coordinate (o1) at (0.0cm,3.75cm);
      \coordinate (o2) at (1.5cm,3.75cm);

      \coordinate (v1) at (0,+1.5cm);
      \coordinate (v2) at (0,+3cm);
      \coordinate (v3) at (1.5cm,+1.5cm);
      \coordinate (v4) at (1.5cm,+3cm);

      \draw [scalar] (i1) -- (v1) node [pos=0.5,anchor=east] {\footnotesize $\frac{1}{N_{\smash[b]{\mathrlap{\!f}}}}$};
      \draw [scalar] (i2) -- (v3) node [pos=0.5,anchor=west] {\footnotesize $\!\frac{1}{N_{\smash[b]{\mathrlap{\!f}}}}$};

      \draw [scalar] (v2) to (o1);
      \draw [scalar] (v4) to (o2);

      \draw [hyp fermion] (v2) -- (v1) -- (v3) -- (v4) -- cycle;

      \draw[decorate,arrow={0.55}{gray}] ($(i2)+(135:0.15cm)$) -- ($(i1)+(45:0.15cm)$);
      \draw[decorate,arrow={0.55}{gray}] ($(v3)+(135:0.15cm)$) -- ($(v1)+(45:0.15cm)$);

      \draw [gray,rounded corners] ($(i1)+(45:0.15cm)$) -- ($(v1)+(315:0.15cm)$) -- ($(v3)+(225:0.15cm)$) -- ($(i2)+(135:0.15cm)$) -- cycle;
      \draw [gray,rounded corners] ($(v1)+(45:0.15cm)$) -- ($(v2)+(315:0.15cm)$) -- ($(v4)+(225:0.15cm)$) -- ($(v3)+(135:0.15cm)$) -- cycle;

      \path (i1) -- (v3) node [pos=0.5] {\footnotesize $N_{\smash[b]{\mathrlap{\!c}}}$};
      \path (v1) -- (v4) node [pos=0.5] {\footnotesize $N_{\smash[b]{\mathrlap{\!f}}}$};

      \fill [] (v1) circle [radius=0.0625cm];
      \fill [] (v2) circle [radius=0.0625cm];
      \fill [] (v3) circle [radius=0.0625cm];
      \fill [] (v4) circle [radius=0.0625cm];

      \draw [thick,rounded corners=2pt,fill=gray] ($(i1)+(-0.25cm,-0.125cm)$) -- ($(i1)+(-0.25cm,0)$) -- ($(i2)+(0.25cm,0)$) -- ($(i2)+(0.25cm,-0.125cm)$) -- cycle;

%      \node at ($(i1)!0.5!(i2)+(0,-0.2cm)$) [anchor=north] {\small $N_c/N_f$};
    \end{tikzpicture}
  }
  \qquad
  \subfloat[$N_c/N_f$]{
    \begin{tikzpicture}
      \coordinate (i1) at (0.0cm,0.0cm);
      \coordinate (i2) at (1.5cm,0.0cm);

      \coordinate (o1) at (0.0cm,3.75cm);
      \coordinate (o2) at (1.5cm,3.75cm);

      \coordinate (v1) at (0.0cm, +0.75cm);
      \coordinate (v2) at (1.5cm, +0.75cm);
      \coordinate (v3) at (0.75cm,+1.75cm);

      \coordinate (v4) at (0.0cm, +3.25cm);
      \coordinate (v5) at (1.5cm, +3.25cm);
      \coordinate (v6) at (0.75cm,+2.25cm);

      \draw [scalar] (i1) -- (v1) node [pos=0.5,anchor=east] {\footnotesize $\frac{1}{N_{\smash[b]{\mathrlap{\!f}}}}$};
      \draw [scalar] (i2) -- (v2) node [pos=0.5,anchor=west] {\footnotesize $\!\frac{1}{N_{\smash[b]{\mathrlap{\!f}}}}$};;

      \draw [scalar] (v4) -- (o1);
      \draw [scalar] (v5) -- (o2);

      \draw [gluon] (v3) -- (v6) node [pos=0.5,anchor=east] {\footnotesize $\frac{1}{N_{\smash[b]{\mathrlap{\!f}}}}$};

      \draw [hyp fermion] (v1) -- (v2) -- (v3) -- cycle;
      \draw [hyp fermion] (v4) -- (v5) -- (v6) -- cycle;

      \draw[decorate,arrow={0.55}{gray}] ($(i2)+(135:0.15cm)$) -- ($(i1)+(45:0.15cm)$);
      \draw[decorate,arrow={0.55}{gray}] ($(v2)+(150:0.15cm)$) -- ($(v1)+(30:0.15cm)$);
      \draw[decorate,arrow={0.55}{gray}] ($(v4)+(330:0.15cm)$) -- ($(v5)+(210:0.15cm)$);

      \draw [gray,rounded corners] ($(i1)+(45:0.15cm)$) -- ($(v1)+(315:0.15cm)$) -- ($(v2)+(225:0.15cm)$) -- ($(i2)+(135:0.15cm)$) -- cycle;
      \draw [gray,rounded corners] ($(v1)+(30:0.15cm)$) -- ($(v2)+(150:0.15cm)$) -- ($(v3)+(270:0.15cm)$) -- cycle;
      \draw [gray,rounded corners] ($(v4)+(330:0.15cm)$) -- ($(v5)+(210:0.15cm)$) -- ($(v6)+(90:0.15cm)$) -- cycle;

      \path  (i1) -- (v2) node [pos=0.5] {\footnotesize $N_{\smash[b]{\mathrlap{\!c}}}$};

      \path  (v3) -- ($(v1)!0.5!(v2)$) node [pos=0.6] {\footnotesize $N_{\smash[b]{\mathrlap{\!f}}}$};
      \path  (v6) -- ($(v4)!0.5!(v5)$) node [pos=0.6] {\footnotesize $N_{\smash[b]{\mathrlap{\!f}}}$};

      \fill [] (v1) circle [radius=0.0625cm];
      \fill [] (v2) circle [radius=0.0625cm];
      \fill [] (v3) circle [radius=0.0625cm];
      \fill [] (v4) circle [radius=0.0625cm];
      \fill [] (v5) circle [radius=0.0625cm];
      \fill [] (v6) circle [radius=0.0625cm];

      \draw [thick,rounded corners=2pt,fill=gray] ($(i1)+(-0.25cm,-0.125cm)$) -- ($(i1)+(-0.25cm,0)$) -- ($(i2)+(0.25cm,0)$) -- ($(i2)+(0.25cm,-0.125cm)$) -- cycle;

%      \node at ($(i1)!0.5!(i2)+(0,-0.2cm)$) [anchor=north] {\small $N_c/N_f$};
    \end{tikzpicture}
  }
  \qquad
  \subfloat[$(N_c/N_f)^2$]{
    \begin{tikzpicture}

      \coordinate (i1) at (0.0cm,0.0cm);
      \coordinate (i2) at (1.5cm,0.0cm);

      \coordinate (o1) at (0.0cm,3.75cm);
      \coordinate (o2) at (1.5cm,3.75cm);

      \coordinate (v1) at (0,+1.5cm);
      \coordinate (v2) at (0,+3cm);
      \coordinate (v3) at (1.5cm,+1.5cm);
      \coordinate (v4) at (1.5cm,+3cm);

      \draw [scalar] (i1) -- (v1) node [pos=0.5,anchor=east] {\footnotesize $\frac{1}{N_{\smash[b]{\mathrlap{\!f}}}}$};
      \draw [scalar] (i2) -- (v3) node [pos=0.5,anchor=west] {\footnotesize $\!\frac{1}{N_{\smash[b]{\mathrlap{\!f}}}}$};

      \draw [scalar] (v2) to (o1);
      \draw [scalar] (v4) to (o2);

      \draw [adj fermion] (v2) -- (v1) -- (v3) -- (v4) -- cycle;

      \draw [decorate,arrow={0.55}{gray}] ($(i2)+(135:0.15cm)$) -- ($(i1)+(45:0.15cm)$);
      \draw [decorate,arrow={0.55}{gray}] ($(v3)+(135:0.15cm)$) -- ($(v1)+(45:0.15cm)$); 

      \draw [gray,rounded corners] ($(i1)+(45:0.15cm)$) -- ($(v1)+(315:0.15cm)$) -- ($(v3)+(225:0.15cm)$) -- ($(i2)+(135:0.15cm)$) -- cycle;
      \draw [gray,rounded corners] ($(v1)+(45:0.15cm)$) -- ($(v2)+(315:0.15cm)$) -- ($(v4)+(225:0.15cm)$) -- ($(v3)+(135:0.15cm)$) -- cycle;

      \path (i1) -- (v3) node [pos=0.5] {\footnotesize $N_{\smash[b]{\mathrlap{\!c}}}$};
      \path (v1) -- (v4) node [pos=0.5] {\footnotesize $N_{\smash[b]{\mathrlap{\!c}}}$};

      \fill [] (v1) circle [radius=0.0625cm];
      \fill [] (v2) circle [radius=0.0625cm];
      \fill [] (v3) circle [radius=0.0625cm];
      \fill [] (v4) circle [radius=0.0625cm];

      \draw [thick,rounded corners=2pt,fill=gray] ($(i1)+(-0.25cm,-0.125cm)$) -- ($(i1)+(-0.25cm,0)$) -- ($(i2)+(0.25cm,0)$) -- ($(i2)+(0.25cm,-0.125cm)$) -- cycle;

%      \node at ($(i1)!0.5!(i2)+(0,-0.2cm)$) [anchor=north] {\small $(N_c/N_f)^{\smash[b]{\mathrlap{2}}}$};
    \end{tikzpicture}
  }

  \caption{Example of counting of factors of $N_c$ and $N_f$ for three diagrams. The double lines represent adjoint fields and the single line give particles transforming in the fundamental representation of the gauge group. Each closed loop gives a factor of $N_c$ or $N_f$ depending on which representation the fields in the loop transform under. Additionally, each propagator for a field in the vector multiplet give a factor $1/N_f$. The first two diagrams are proportional to $N_c/N_f$ and hence count as one-loop diagrams. The dotted propagator in the third diagram corresponds to a fermion in the adjoint hypermultiplet. This diagram is proportional to $(N_c/N_f)^2$ and is therefore suppressed in the weakly-coupled planar limit.}
  \label{fig:Nf-Nc-counting}
\end{figure}
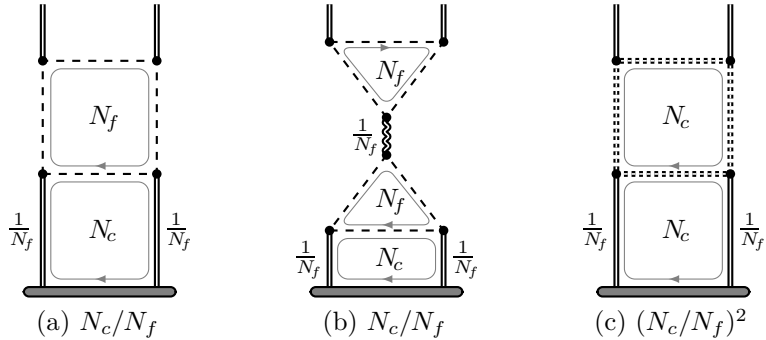

In principle we should also consider the diagrams above, but with the fields from the fundamental hypermultiplet replaced by the corresponding fields in the adjoint hypermultiplet. However, in that case the sums over flavours in the loops are replaced by a sum over colours. Hence, the resulting diagrams are suppressed at small $\lambda$. The counting of factors of $N_c$ and $N_f$ is illustrated in figure~\ref{fig:Nf-Nc-counting}.

Let us now look at the flavour structure of the diagrams in~\eqref{eq:D-one-loop-diagrams}. The second and fourth diagram give no non-trivial $\algSO(4)$ flavour interactions. As discussed above, the coefficient of the $\algSO(4)$ identity tensor $\mathds{1}$ in the dilatation operator can be determined using supersymmetry. Hence, we do not need to calculate those diagrams. The third diagram in~\eqref{eq:D-one-loop-diagrams} gives an $\algSO(4)$ trace in flavour space. By expanding the effective vertices in this diagram in terms of the one-loop diagrams with a fundamental hyper, we find four different diagrams. However, all four vanish due to symmetry as detailed in appendix~\ref{app:feynman}. Hence, the first diagram in~\eqref{eq:D-one-loop-diagrams} is the only one we have to compute.

\subsection{Computation of one-loop dilatation operator}

In this sub-section we consider the first diagram in~\eqref{eq:D-one-loop-diagrams}. 
As was shown above, it is only this diagram that has a potentially non-trivial $\algSO(4)$ flavour structure and hence is needed to determine the dilatation operator. We may expand the diagram by plugging in the effective vertices~\eqref{eq:effvertex1} and~\eqref{eq:effvertex2} 
\begin{equation}
  \begin{tikzpicture}[baseline=-0.5ex]

    \coordinate (i1) at (0,-0.75cm);
    \coordinate (i2) at (1cm,-0.75cm);

    \coordinate (o1) at (0,0.75cm);
    \coordinate (o2) at (1cm,0.75cm);

    \coordinate (v) at (0.5cm,0cm);

    \draw [scalar,out=90,in=225] (i1) to (v);
    \draw [scalar,out=90,in=315] (i2) to (v);

    \draw [scalar,out=135,in=270] (v) to (o1);
    \draw [scalar,out=45,in=270] (v) to (o2);

    \fill [] (v) circle [radius=0.125cm];

    \draw [thick,rounded corners=2pt,fill=gray] ($(i1)+(-0.25cm,-0.125cm)$) -- ($(i1)+(-0.25cm,0)$) -- ($(i2)+(0.25cm,0)$) -- ($(i2)+(0.25cm,-0.125cm)$) -- cycle;
  \end{tikzpicture}
 \quad =\quad
    \begin{tikzpicture}[baseline=-0.5ex]
    \coordinate (i1) at (0,-0.75cm);
    \coordinate (i2) at (1cm,-0.75cm);

    \coordinate (o1) at (0,0.75cm);
    \coordinate (o2) at (1cm,0.75cm);

    \coordinate (v1) at (0,0);
    \coordinate (v2) at (1cm,0);

    \draw [scalar] (i1) to (v1);
    \draw [scalar] (i2) to (v2);

    \draw [scalar] (v1) to (o1);
    \draw [scalar] (v2) to (o2);

    \draw [hyp scalar,out=45,in=135] (v1) to (v2);
    \draw [hyp scalar,out=315,in=225] (v1) to (v2);

    \fill [] (v1) circle [radius=0.0625cm];
    \fill [] (v2) circle [radius=0.0625cm];

    \draw [thick,rounded corners=2pt,fill=gray] ($(i1)+(-0.25cm,-0.125cm)$) -- ($(i1)+(-0.25cm,0)$) -- ($(i2)+(0.25cm,0)$) -- ($(i2)+(0.25cm,-0.125cm)$) -- cycle;
  \end{tikzpicture}
  \quad +\quad
  \begin{tikzpicture}[baseline=-0.5ex]
    \coordinate (i1) at (0,-0.75cm);
    \coordinate (i2) at (1cm,-0.75cm);

    \coordinate (o1) at (0,0.75cm);
    \coordinate (o2) at (1cm,0.75cm);

    \coordinate (v1) at (0.5cm,-0.4cm);
    \coordinate (v2) at (0.5cm,+0.4cm);

    \draw [scalar,out=90,in=225] (i1) to (v1);
    \draw [scalar,out=90,in=315] (i2) to (v1);

    \draw [scalar,out=135,in=270] (v2) to (o1);
    \draw [scalar,out=45,in=270] (v2) to (o2);

    \draw [hyp scalar,out=45,in=315,looseness=1.5] (v1) to (v2);
    \draw [hyp scalar,out=135,in=225,looseness=1.5] (v1) to (v2);

    \fill [] (v1) circle [radius=0.0625cm];
    \fill [] (v2) circle [radius=0.0625cm];

    \draw [thick,rounded corners=2pt,fill=gray] ($(i1)+(-0.25cm,-0.125cm)$) -- ($(i1)+(-0.25cm,0)$) -- ($(i2)+(0.25cm,0)$) -- ($(i2)+(0.25cm,-0.125cm)$) -- cycle;
  \end{tikzpicture}
  \quad + \quad
    \begin{tikzpicture}[baseline=-0.5ex]
    \coordinate (i1) at (0,-0.75cm);
    \coordinate (i2) at (1cm,-0.75cm);

    \coordinate (o1) at (0,0.75cm);
    \coordinate (o2) at (1cm,0.75cm);

    \coordinate (v1) at (0,-0.3cm);
    \coordinate (v2) at (0,+0.3cm);
    \coordinate (v3) at (1cm,-0.3cm);
    \coordinate (v4) at (1cm,+0.3cm);

    \draw [scalar] (i1) to (v1);
    \draw [scalar] (i2) to (v3);

    \draw [scalar] (v2) to (o1);
    \draw [scalar] (v4) to (o2);

    \draw [hyp fermion] (v2) to (v1) to (v3) to (v4) to (v2);

    \fill [] (v1) circle [radius=0.0625cm];
    \fill [] (v2) circle [radius=0.0625cm];
    \fill [] (v3) circle [radius=0.0625cm];
    \fill [] (v4) circle [radius=0.0625cm];

    \draw [thick,rounded corners=2pt,fill=gray] ($(i1)+(-0.25cm,-0.125cm)$) -- ($(i1)+(-0.25cm,0)$) -- ($(i2)+(0.25cm,0)$) -- ($(i2)+(0.25cm,-0.125cm)$) -- cycle;
  \end{tikzpicture}\,.
\end{equation}
In fact, the first of these diagram again has a trivial flavour structure and so we do not need to calculate it. The second diagram gives an $\algSO(4)$ trace in flavour space, and we compute it in appendix~\ref{app:feynman}. There we show that the corresponding integral is UV-finite, and therefore does not contribute to the mixing matrix. We are only left with the last diagram, the fermion box
\begin{equation}\label{eq:diagram-fermion-square}
  \begin{tikzpicture}[baseline=-0.5ex]
    \coordinate (i1) at (0,-0.75cm);
    \coordinate (i2) at (1cm,-0.75cm);

    \coordinate (o1) at (0,0.75cm);
    \coordinate (o2) at (1cm,0.75cm);

    \coordinate (v1) at (0,-0.3cm);
    \coordinate (v2) at (0,+0.3cm);
    \coordinate (v3) at (1cm,-0.3cm);
    \coordinate (v4) at (1cm,+0.3cm);

    \draw [scalar] (i1) to (v1);
    \draw [scalar] (i2) to (v3);

    \draw [scalar] (v2) to (o1);
    \draw [scalar] (v4) to (o2);

    \draw [hyp fermion] (v2) to (v1) to (v3) to (v4) to (v2);

    \fill [] (v1) circle [radius=0.0625cm];
    \fill [] (v2) circle [radius=0.0625cm];
    \fill [] (v3) circle [radius=0.0625cm];
    \fill [] (v4) circle [radius=0.0625cm];

    \draw [thick,rounded corners=2pt,fill=gray] ($(i1)+(-0.25cm,-0.125cm)$) -- ($(i1)+(-0.25cm,0)$) -- ($(i2)+(0.25cm,0)$) -- ($(i2)+(0.25cm,-0.125cm)$) -- cycle;
  \end{tikzpicture}\,.
\end{equation}
The left- and right-moving fermions $\lambda_{\sL}$ and $\lambda_{\sR}$ in the hypermultiplet are charged under $\algSU(2)_{\sL}$ and $\algSU(2)_{\sR}$, respectively. The cubic vertices in the fermion box diagram in~\eqref{eq:diagram-fermion-square} each couple $\phi$ to $\lambda_{\sL}$ and $\lambda_{\sR}$. Hence, the diagram in~\eqref{eq:diagram-fermion-square} in fact corresponds to two diagrams that differ in the assignment of chiralities of the fermions in the loop. It turns out that  the two diagrams result in two different flavour structures.

\begin{figure}
  \centering
  
  \subfloat[$\delta_{a_{k\protect\vphantom{+}}}^{b_{k\protect\vphantom{+}}} \delta_{a_{k+1}}^{b_{k+1}} \epsilon_{\dot{a}_{k} \dot{a}_{k+1}} \epsilon^{\dot{b}_{k} \dot{b}_{k+1}}$]{
    \begin{tikzpicture}[
      baseline=-0.5ex,
      suL/.style={thick,red,dash pattern=on 2mm off 0.4mm,rounded corners},
      suR/.style={thick,blue,rounded corners}
      ]

      \begin{scope}[xshift=-0.75cm-1.5cm,yshift=-1.5cm]
        \coordinate (i1) at (0.0cm,0.0cm);
        \coordinate (i2) at (1.5cm,0.0cm);

        \coordinate (o1) at (0.0cm,3.0cm);
        \coordinate (o2) at (1.5cm,3.0cm);

        \coordinate (v1) at (0,+1.25cm);
        \coordinate (v2) at (0,+2.50cm);
        \coordinate (v3) at (1.5cm,+1.25cm);
        \coordinate (v4) at (1.5cm,+2.50cm);

        \useasboundingbox ($(i1)+(-0.25cm,-0.5cm)$) rectangle ($(o2)+(+0.25cm,+0.25cm)$);

        \draw [scalar] (i1) -- (v1);
        \draw [scalar] (i2) -- (v3);

        \draw [scalar] (v2) to (o1);
        \draw [scalar] (v4) to (o2);

        \draw [hyp fermion] (v2) -- (v1) -- (v3) -- (v4) -- cycle;

        \fill [] (v1) circle [radius=0.0625cm];
        \fill [] (v2) circle [radius=0.0625cm];
        \fill [] (v3) circle [radius=0.0625cm];
        \fill [] (v4) circle [radius=0.0625cm];

        \node at (v1) [anchor=south east,outer sep=0,inner sep=2pt] {\scriptsize $\bar{\lambda}_{\sL}$};
        \node at (v1) [anchor=north east,outer sep=0,inner sep=2pt] {\scriptsize $\phi$};
        \node at (v1) [anchor=north west,outer sep=0,inner sep=2pt] {\scriptsize $\mathrlap{\lambda_{\sR}}\phantom{\bar{\lambda}_{\sL}}$};

        \node at (v2) [anchor=south west,outer sep=0,inner sep=2pt] {\scriptsize $\bar{\lambda}_{\sR}$};
        \node at (v2) [anchor=south east,outer sep=0,inner sep=2pt] {\scriptsize $\phi$};
        \node at (v2) [anchor=north east,outer sep=0,inner sep=2pt] {\scriptsize $\lambda_{\sL}$};

        \node at (v3) [anchor=north east,outer sep=0,inner sep=2pt] {\scriptsize $\bar{\lambda}_{\sR}$};
        \node at (v3) [anchor=north west,outer sep=0,inner sep=2pt] {\scriptsize $\phi$};
        \node at (v3) [anchor=south west,outer sep=0,inner sep=2pt] {\scriptsize $\lambda_{\sL}$};

        \node at (v4) [anchor=north west,outer sep=0,inner sep=2pt] {\scriptsize $\bar{\lambda}_{\sL}$};
        \node at (v4) [anchor=south west,outer sep=0,inner sep=2pt] {\scriptsize $\phi$};
        \node at (v4) [anchor=south east,outer sep=0,inner sep=2pt] {\scriptsize $\lambda_{\sR}$};

        \draw [thick,rounded corners=2pt,fill=gray] ($(i1)+(-0.25cm,-0.125cm)$) -- ($(i1)+(-0.25cm,0)$) -- ($(i2)+(0.25cm,0)$) -- ($(i2)+(0.25cm,-0.125cm)$) -- cycle;
      \end{scope}

      \node at (0,0) {$=$};

      \begin{scope}[xshift=-0.75cm+1.5cm,yshift=-1.5cm]
        \coordinate (i1) at (0.0cm,0.0cm);
        \coordinate (i2) at (1.5cm,0.0cm);

        \coordinate (o1) at (0.0cm,3.0cm);
        \coordinate (o2) at (1.5cm,3.0cm);

        \coordinate (v1) at (0,+1.25cm);
        \coordinate (v2) at (0,+2.50cm);
        \coordinate (v3) at (1.5cm,+1.25cm);
        \coordinate (v4) at (1.5cm,+2.50cm);

        \draw [lightgray,scalar] (i1) -- (v1);
        \draw [lightgray,scalar] (i2) -- (v3);

        \draw [lightgray,scalar] (v2) to (o1);
        \draw [lightgray,scalar] (v4) to (o2);

        \draw [lightgray,hyp fermion] (v2) -- (v1) -- (v3) -- (v4) -- cycle;

        \fill [lightgray] (v1) circle [radius=0.0625cm];
        \fill [lightgray] (v2) circle [radius=0.0625cm];
        \fill [lightgray] (v3) circle [radius=0.0625cm];
        \fill [lightgray] (v4) circle [radius=0.0625cm];

        \draw [suR] ($(i1)+(+0.1cm,0)$) -- ($(v1)+(+0.1cm,-0.08cm)$) -- ($(v3)+(+0.1cm,-0.08cm)$) -- ($(i2)+(+0.1cm,0)$);

        \draw [decorate,arrow={0.55}{blue}] ($(i1)+(+0.1cm,0)$) -- ($(v1)+(+0.1cm,-0.08cm)$);
        \draw [decorate,arrow={0.50}{blue}] ($(v1)+(+0.1cm,-0.08cm)$) -- ($(v3)+(+0.1cm,-0.08cm)$);
        \draw [decorate,arrow={0.60}{blue}] ($(v3)+(+0.1cm,-0.08cm)$) -- ($(i2)+(+0.1cm,0)$);

        \draw [suR] ($(o2)+(+0.1cm,0)$) -- ($(v4)+(+0.1cm,+0.08cm)$) -- ($(v2)+(+0.1cm,+0.08cm)$) -- ($(o1)+(+0.1cm,0)$);

        \draw [decorate,arrow={0.66}{blue}] ($(o2)+(+0.1cm,0)$) -- ($(v4)+(+0.1cm,+0.08cm)$);
        \draw [decorate,arrow={0.65}{blue}] ($(v4)+(+0.1cm,+0.08cm)$) -- ($(v2)+(+0.1cm,+0.08cm)$);
        \draw [decorate,arrow={0.66}{blue}] ($(v2)+(+0.1cm,+0.08cm)$) -- ($(o1)+(+0.1cm,0)$);

        \draw [suL] ($(o1)+(-0.1cm,0)$) -- ($(i1)+(-0.1cm,0)$);
        \draw [decorate,arrow={0.66}{red}] ($(v1)+(-0.1cm,-0.08cm)$) -- ($(i1)+(-0.1cm,0)$);
        \draw [decorate,arrow={0.66}{red}] ($(v2)+(-0.1cm,+0.08cm)$) -- ($(v1)+(-0.1cm,-0.08cm)$);
        \draw [decorate,arrow={0.66}{red}] ($(o1)+(-0.1cm,0)$) -- ($(v2)+(-0.1cm,+0.08cm)$);

        \draw [suL] ($(o2)+(-0.1cm,0)$) -- ($(i2)+(-0.1cm,0)$);
        \draw [decorate,arrow={0.66}{red}] ($(i2)+(-0.1cm,0)$) -- ($(v3)+(-0.1cm,-0.08cm)$);
        \draw [decorate,arrow={0.66}{red}] ($(v3)+(-0.1cm,+0.08cm)$) -- ($(v4)+(-0.1cm,-0.08cm)$);
        \draw [decorate,arrow={0.66}{red}] ($(v4)+(-0.1cm,+0.08cm)$) -- ($(o2)+(-0.1cm,0)$);

        \node at (i1) [anchor=south east,outer sep=0,inner sep=2pt] {\scriptsize $a_{k\vphantom{+}}\,\,$};
        \node at (i1) [anchor=south west ,outer sep=0,inner sep=2pt] {\scriptsize $\,\,\dot{a}_{k\vphantom{+}}$};

        \node at (i2) [anchor=south east,outer sep=0,inner sep=2pt] {\scriptsize $a_{k+1}\,\,$};
        \node at (i2) [anchor=south west ,outer sep=0,inner sep=2pt] {\scriptsize $\,\,\dot{a}_{k+1}$};

        \node at (o1) [anchor=east,outer sep=0,inner sep=2pt] {\scriptsize $b_{k\vphantom{+}}\,\,$};
        \node at (o1) [anchor=west ,outer sep=0,inner sep=2pt] {\scriptsize $\,\,\dot{b}_{k\vphantom{+}}$};

        \node at (o2) [anchor=east,outer sep=0,inner sep=2pt] {\scriptsize $b_{k+1}\,\,$};
        \node at (o2) [anchor=west ,outer sep=0,inner sep=2pt] {\scriptsize $\,\,\dot{b}_{k+1}$};

        \draw [thick,rounded corners=2pt,fill=gray] ($(i1)+(-0.25cm,-0.125cm)$) -- ($(i1)+(-0.25cm,0)$) -- ($(i2)+(0.25cm,0)$) -- ($(i2)+(0.25cm,-0.125cm)$) -- cycle;

%        \path (i1) -- (i2) node [pos=0.5,anchor=north,outer sep=8pt] {\scriptsize $\delta_{a_k}^{b_k} \delta_{a_{k+1}}^{b_{k+1}} \epsilon_{\dot{a}_{k} \dot{a}_{k+1}} \epsilon^{\dot{b}_{k} \dot{b}_{k+1}}$}; %
      \end{scope}
    \end{tikzpicture}
  }%
  \qquad
  \subfloat[$\epsilon_{a_{k} a_{k+1}} \epsilon^{b_{k} b_{k+1}} \delta_{\dot{a}_{k\protect\vphantom{+}}}^{\dot{b}_{k\protect\vphantom{+}}} \delta_{\dot{a}_{k+1}}^{\dot{b}_{k+1}}$]{
    \begin{tikzpicture}[
      baseline=-0.5ex,
      suL/.style={thick,red,dash pattern=on 2mm off 0.4mm,rounded corners},
      suR/.style={thick,blue,rounded corners}
      ]

      % \begin{scope}[xshift=+3.5cm]
      \begin{scope}[xshift=-0.75cm-1.5cm,yshift=-1.5cm]
        \coordinate (i1) at (0.0cm,0.0cm);
        \coordinate (i2) at (1.5cm,0.0cm);

        \coordinate (o1) at (0.0cm,3.0cm);
        \coordinate (o2) at (1.5cm,3.0cm);

        \coordinate (v1) at (0,+1.25cm);
        \coordinate (v2) at (0,+2.50cm);
        \coordinate (v3) at (1.5cm,+1.25cm);
        \coordinate (v4) at (1.5cm,+2.50cm);

        \useasboundingbox ($(i1)+(-0.25cm,-0.5cm)$) rectangle ($(o2)+(+0.25cm,+0.25cm)$);

        \draw [scalar] (i1) -- (v1);
        \draw [scalar] (i2) -- (v3);

        \draw [scalar] (v2) to (o1);
        \draw [scalar] (v4) to (o2);

        \draw [hyp fermion] (v2) -- (v1) -- (v3) -- (v4) -- cycle;

        \fill [] (v1) circle [radius=0.0625cm];
        \fill [] (v2) circle [radius=0.0625cm];
        \fill [] (v3) circle [radius=0.0625cm];
        \fill [] (v4) circle [radius=0.0625cm];

        \node at (v1) [anchor=south east,outer sep=0,inner sep=2pt] {\scriptsize $\bar{\lambda}_{\sR}$};
        \node at (v1) [anchor=north east,outer sep=0,inner sep=2pt] {\scriptsize $\phi$};
        \node at (v1) [anchor=north west,outer sep=0,inner sep=2pt] {\scriptsize $\mathrlap{\lambda_{\sL}}\phantom{\bar{\lambda}_{\sR}}$};

        \node at (v2) [anchor=south west,outer sep=0,inner sep=2pt] {\scriptsize $\bar{\lambda}_{\sL}$};
        \node at (v2) [anchor=south east,outer sep=0,inner sep=2pt] {\scriptsize $\phi$};
        \node at (v2) [anchor=north east,outer sep=0,inner sep=2pt] {\scriptsize $\lambda_{\sR}$};

        \node at (v3) [anchor=north east,outer sep=0,inner sep=2pt] {\scriptsize $\bar{\lambda}_{\sL}$};
        \node at (v3) [anchor=north west,outer sep=0,inner sep=2pt] {\scriptsize $\phi$};
        \node at (v3) [anchor=south west,outer sep=0,inner sep=2pt] {\scriptsize $\lambda_{\sR}$};

        \node at (v4) [anchor=north west,outer sep=0,inner sep=2pt] {\scriptsize $\bar{\lambda}_{\sR}$};
        \node at (v4) [anchor=south west,outer sep=0,inner sep=2pt] {\scriptsize $\phi$};
        \node at (v4) [anchor=south east,outer sep=0,inner sep=2pt] {\scriptsize $\lambda_{\sL}$};

        \draw [thick,rounded corners=2pt,fill=gray] ($(i1)+(-0.25cm,-0.125cm)$) -- ($(i1)+(-0.25cm,0)$) -- ($(i2)+(0.25cm,0)$) -- ($(i2)+(0.25cm,-0.125cm)$) -- cycle;
      \end{scope}

      \node at (0,0) {$=$};

      \begin{scope}[xshift=-0.75cm+1.5cm,yshift=-1.5cm]
        \coordinate (i1) at (0.0cm,0.0cm);
        \coordinate (i2) at (1.5cm,0.0cm);

        \coordinate (o1) at (0.0cm,3.0cm);
        \coordinate (o2) at (1.5cm,3.0cm);

        \coordinate (v1) at (0,+1.25cm);
        \coordinate (v2) at (0,+2.50cm);
        \coordinate (v3) at (1.5cm,+1.25cm);
        \coordinate (v4) at (1.5cm,+2.50cm);

        \draw [lightgray,scalar] (i1) -- (v1);
        \draw [lightgray,scalar] (i2) -- (v3);

        \draw [lightgray,scalar] (v2) to (o1);
        \draw [lightgray,scalar] (v4) to (o2);

        \draw [lightgray,hyp fermion] (v2) -- (v1) -- (v3) -- (v4) -- cycle;

        \fill [lightgray] (v1) circle [radius=0.0625cm];
        \fill [lightgray] (v2) circle [radius=0.0625cm];
        \fill [lightgray] (v3) circle [radius=0.0625cm];
        \fill [lightgray] (v4) circle [radius=0.0625cm];

        \draw [suR] ($(i1)+(+0.1cm,0)$) -- ($(o1)+(+0.1cm,0)$);
        \draw [decorate,arrow={0.66}{blue}] ($(i1)+(+0.1cm,0)$) -- ($(v1)+(+0.1cm,-0.08cm)$);
        \draw [decorate,arrow={0.66}{blue}] ($(v1)+(+0.1cm,+0.08cm)$) -- ($(v2)+(+0.1cm,-0.08cm)$);
        \draw [decorate,arrow={0.66}{blue}] ($(v2)+(+0.1cm,+0.08cm)$) -- ($(o1)+(+0.1cm,0)$);

        \draw [suR] ($(o2)+(+0.1cm,0)$) -- ($(i2)+(+0.1cm,0)$);
        \draw [decorate,arrow={0.66}{blue}] ($(v3)+(+0.1cm,-0.08cm)$) -- ($(i2)+(+0.1cm,0)$);
        \draw [decorate,arrow={0.66}{blue}] ($(v4)+(+0.1cm,-0.08cm)$) -- ($(v3)+(+0.1cm,+0.08cm)$);
        \draw [decorate,arrow={0.66}{blue}] ($(o2)+(+0.1cm,0)$) -- ($(v4)+(+0.1cm,+0.08cm)$);

        \draw [suL] ($(i2)+(-0.1cm,0)$) -- ($(v3)+(-0.1cm,-0.08cm)$) -- ($(v1)+(-0.1cm,-0.08cm)$) -- ($(i1)+(-0.1cm,0)$);
        \draw [decorate,arrow={0.66}{red}] ($(v1)+(-0.1cm,-0.08cm)$) -- ($(i1)+(-0.1cm,0)$);
        \draw [decorate,arrow={0.60}{red}] ($(v3)+(-0.1cm,-0.08cm)$) -- ($(v1)+(-0.1cm,-0.08cm)$);
        \draw [decorate,arrow={0.66}{red}] ($(i2)+(-0.1cm,0)$) -- ($(v3)+(-0.1cm,-0.08cm)$);

        \draw [suL] ($(o2)+(-0.1cm,0)$) -- ($(v4)+(-0.1cm,+0.08cm)$) -- ($(v2)+(-0.1cm,+0.08cm)$) -- ($(o1)+(-0.1cm,0)$);
        \draw [decorate,arrow={0.66}{red}] ($(v4)+(-0.1cm,+0.08cm)$) -- ($(o2)+(-0.1cm,0)$);
        \draw [decorate,arrow={0.60}{red}] ($(v2)+(-0.1cm,+0.08cm)$) -- ($(v4)+(-0.1cm,+0.08cm)$);
        \draw [decorate,arrow={0.66}{red}] ($(o1)+(-0.1cm,0)$) -- ($(v2)+(-0.1cm,+0.08cm)$);

        \node at (i1) [anchor=south east,outer sep=0,inner sep=2pt] {\scriptsize $a_{k\vphantom{+}}\,\,$};
        \node at (i1) [anchor=south west ,outer sep=0,inner sep=2pt] {\scriptsize $\,\,\dot{a}_{k\vphantom{+}}$};

        \node at (i2) [anchor=south east,outer sep=0,inner sep=2pt] {\scriptsize $a_{k+1}\,\,$};
        \node at (i2) [anchor=south west ,outer sep=0,inner sep=2pt] {\scriptsize $\,\,\dot{a}_{k+1}$};

        \node at (o1) [anchor=east,outer sep=0,inner sep=2pt] {\scriptsize $b_{k\vphantom{+}}\,\,$};
        \node at (o1) [anchor=west ,outer sep=0,inner sep=2pt] {\scriptsize $\,\,\dot{b}_{k\vphantom{+}}$};

        \node at (o2) [anchor=east,outer sep=0,inner sep=2pt] {\scriptsize $b_{k+1}\,\,$};
        \node at (o2) [anchor=west ,outer sep=0,inner sep=2pt] {\scriptsize $\,\,\dot{b}_{k+1}$};

        \draw [thick,rounded corners=2pt,fill=gray] ($(i1)+(-0.25cm,-0.125cm)$) -- ($(i1)+(-0.25cm,0)$) -- ($(i2)+(0.25cm,0)$) -- ($(i2)+(0.25cm,-0.125cm)$) -- cycle;

        % \path (i1) -- (i2) node [pos=0.5,anchor=north,outer sep=8pt] {\scriptsize $\epsilon_{a_{k} a_{k+1}} \epsilon^{b_{k} b_{k+1}} \delta_{\dot{a}_k}^{\dot{b}_k} \delta_{\dot{a}_{k+1}}^{\dot{b}_{k+1}}$}; %
      \end{scope}      
      % \end{scope}
    \end{tikzpicture}
  }

  \caption{The flavour structure of the fermion box diagram for the two fermion chiralities. The dashed red lines represent the $\algSU(2)_{\sL}$ flavour, while the $\algSU(2)_{\sR}$ flavour is represented by solid blue lines. A vertical line connecting the operator at the bottom with an external line corresponds to a Kronecker $\delta$ while a line connecting two sites of the operator or two external lines corresponds to a Levi-Civita tensor $\epsilon$.}
  \label{fig:fermion-box-flavour-struct}
\end{figure}
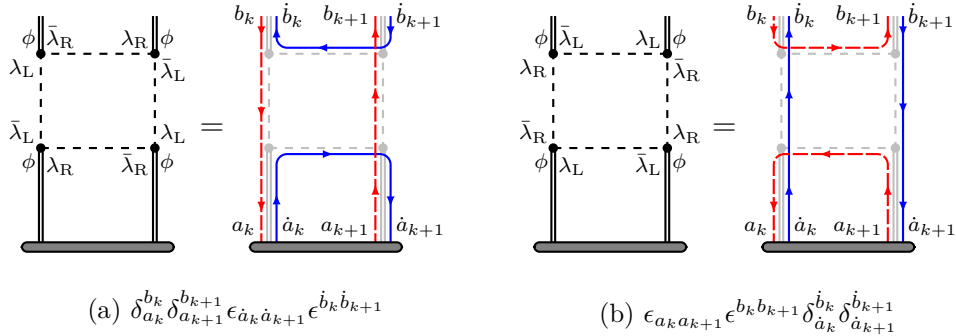

Let us consider such flavour structures in more detail. In each propagator and each vertex the $\algSU(2)_{\sL}$ and $\algSU(2)_{\sR}$ indices are contracted by invariant $\epsilon$-symbols as shown in figure~\ref{fig:fermion-box-flavour-struct}. In the figure the $\algSU(2)_{\sL}$ ($\algSU(2)_{\sR}$) indices are represented by dashed red (solid blue) lines. A line connecting the operator at the bottom with an external line gives a contraction with a Kronecker $\delta$ while a line connecting two sites of the operator or two external lines contracts the indices with a Levi-Civita tensor $\epsilon$. 

The computation of the divergence for each of these diagrams is somewhat involved due to the propagator~\eqref{eq:phi-two-pt-fn}. It is straightforward, however, to show that the \emph{difference} of the two resulting integrals is finite, see appendix~\ref{app:feynman}. To perform the calculation for the \emph{sum} of the diagrams we work in two steps. We first compute the integral in dimension~$d$ using the propagator of~\eqref{eq:phi-two-pt-fn}, that in momentum space is
\begin{equation}
\label{eq:phi-two-pt-fn-fourier}
\braket{\phi^{\alpha\dot{\alpha}}(-p) \phi^{\beta\dot{\beta}}(+p)} = \frac{C' \epsilon^{\alpha\beta} \epsilon^{\dot{\alpha}\dot{\beta}}}{|p|^{2(d/2-1)}}.
\end{equation}
The resulting integral has a UV log-divergence \emph{in any} $d>2$. We then check that the diagram remains UV divergent also when taking $d\to2$. As the diagram~\eqref{eq:diagram-fermion-square} is logarithmically divergent in the UV it contributes to the dilatation operator. 
We again relegate the details of such calculation to appendix~\ref{app:feynman}. 

We have concluded that the divergent part of the fermion-box diagram yields a contribution to the mixing matrix $M$ of the form
\begin{equation}
  M^{b_k \dot{b}_k b_{k+1} \dot{b}_{k+1}}_{a_k \dot{a}_k a_{k+1} \dot{a}_{k+1}} \propto
  \delta^{b_k}_{a_k} \delta^{b_{k+1}}_{a_{k+1}} \epsilon^{\dot{b}_k \dot{b}_{k+1}} \epsilon_{\dot{a}_k \dot{a}_{k+1}}
  +
  \epsilon^{b_k b_{k+1}} \epsilon_{a_k a_{k+1}} \delta^{\dot{b}_k}_{\dot{a}_k} \delta^{\dot{a}_{k+1}}_{\dot{a}_{k+1}}\,.
\end{equation}
To simplify this we introduce the identity operator $1$ and permutation $P$ acting on the tensor product of two spin-1/2 $\algSU(2)$ representations. If $\Psi^{a}$ is an $\algSU(2)$ doublet this action can be written as
\begin{equation}
  1 \, \ket{\Psi^a \Psi^b} = \ket{\Psi^a \Psi^b} , \qquad
  P \, \ket{\Psi^a \Psi^b} = \ket{\Psi^b \Psi^a} .
\end{equation}
The combination $\frac{1}{2}(1-P)$ then represents a projector onto the $\algSU(2)$ singlet
\begin{equation}
  \tfrac{1}{2} (1-P) \, \ket{\Psi^a \Psi^b} = \tfrac{1}{2} ( \ket{\Psi^a \Psi^b} - \ket{\Psi^b \Psi^a} ) = \tfrac{1}{2} \epsilon^{ab} \epsilon_{cd} \ket{\Psi^c \Psi^d}
\end{equation}
Since the divergence of both fermion box diagrams shown in figure~\ref{fig:fermion-box-flavour-struct} is the same, the flavour structure of the mixing matrix $M$ can then be written as
\begin{equation}
  \begin{aligned}
    M \propto 1 \otimes (1-P) + (1-P) \otimes 1 &= 1 \otimes 1 - P \otimes P + (1-P) \otimes (1-P) \\
    &=\mathds{1}-\mathds{P}+\mathds{K}\,.
  \end{aligned}
\end{equation}
Since the above operator annihilates the supersymmetric vacuum~\eqref{eq:vacuum} by itself, and since the diagrams that we have refrained from computing so far do not, we find that these must cancel by supersymmetry. 

We conclude that the one-loop dilatation operator in this sector is given by
\begin{equation}
\delta\gen{D} \propto \frac{N_c}{N_f} \sum_{k=1}^{L}\big(\mathds{1}-
\mathds{P}+\mathds{K}\big)_{k,k+1}\,,
\end{equation}
in the~$\algSO(4)$ sector, up to an overall numerical constant.
 This is precisely the  Hamiltonian of the integrable $\algSO(4)$ spin-chain~\cite{Reshetikhin:1983vw,Reshetikhin:1986vd}.

\section{Conclusions}
\label{sec:conclusions}
In this paper we have investigated the Higgs-branch $\CFT_2$ states that are localised near the small-instanton singularity. Using Witten's description of such states~\cite{Witten:1997yu}, we showed that they can be analysed in a planar, weak-coupling limit, with $\lambda\equiv N_c/N_f$ playing the role of the 't Hooft coupling and non-planar diagrams suppressed by powers of $1/N_c^2$. We demonstrated that at small $\lambda$ and large $N_c$ computing the anomalous dimension of   such operators reduces to the calculation of the spectrum of a nearest-neighbour spin chain. Further, in a sub-sector of the theory, we showed that the Hamiltonian of the spin-chain is integrable. 

The dynamics of states localised near the origin of the Higgs branch features a vector multiplet and an adjoint-valued hypermultiplet, and is governed by a non-local effective action~\cite{Witten:1997yu}. Remarkably, the field content in this effective theory that is used to describe the $\CFT_2$ spin-chain matches precisely the representations that appear in the local integrable spin-chain obtained by taking the small 't Hooft coupling constant limit of the $\AdS_3$ integrable system~\cite{OhlssonSax:2011ms,Sax:2012jv,Borsato:2013qpa}. For this matching to occur, it is necessary for the auxiliary fields present in the vector multiplet to become dynamical. This property is characteristic of supermultiplets with non-canonical kinetic terms, such as the ones that appear in the action introduced in~\cite{Witten:1997yu}.
Still, the composite operators in this theory are simply the conventional gauge-invariant sequences of adjoint-valued fields, and in the large-$N_c$ limit of the $\CFT_2$ one may restrict to single-trace operators.  We calculated the anomalous dimensions of such operators in the scalar $\algSO(4)$ sub-sector of the theory, a calculation which should be dual to the one of part of the closed-string spectrum. We have shown that this results in an integrable Hamiltonian for the vector representation of $\algSO(4)$ in a manner that is reminiscent of~\cite{Minahan:2002ve}. This is the first  appearance of integrability on the $\CFT_2$ side of the duality. Together with the evidence for integrability that has been found on the string theory side~\cite{Borsato:2014exa,Borsato:2014hja, Lloyd:2014bsa,bossstoappear}, it provides a powerful argument for a holographic integrability description of $\AdS_3/\CFT_2$. This evidence leads us to suggest that the perturbative closed strings of the R-R $\AdS_3$ background are dual to $\CFT_2$ states localised near the small-instanton singularity.

In a forthcoming paper, we will return to the computation of the complete $\CFT_2$  spin-chain Hamiltonian. There, we will also address the issue of length-changing effects that are expected to occur in the complete spin-chain, much as they do in $\mathcal{N}=4$ SYM~\cite{Beisert:2003ys}. This should allow us to understand better, from the weakly-coupled $\CFT_2$ point of view, how the massless modes enter into the reducible spin-chain description introduced in~\cite{Sax:2012jv}. It will also allow for a $\CFT_2$-based understanding of the S matrix constructed 
in~\cite{Borsato:2014exa,Borsato:2014hja}. In this context it will be particularly illuminating to extend the analysis of the chiral-ring of the $\CFT_2$ that was initiated in~\cite{Sax:2012jv}.

It is widely believed that the $\Sym^N(\Torus^4)$ orbifold can be found at some point in the moduli space of the $\mathcal{N}=(4,4)$ $\CFT_2$ that  we have been investigating in this paper. However, it appears to be much harder to understand the spin-chain picture starting from the  $\Sym^N(\Torus^4)$ orbifold point~\cite{Pakman:2009mi}, though many suggestive connections seem to exist. It would be interesting to see whether it is possible to deform the spin-chain~\cite{OhlssonSax:2011ms,Sax:2012jv} to approach the $\Sym^N(\Torus^4)$ orbifold point. It is not clear to us whether the connection to the orbifold can be fully understood in the small $\lambda$ limit.\footnote{We are grateful to Kostya Zarembo for discussions on this point.}

Throughout this paper we have focused on the $\CFT_2$ description of the D1-D5 system, or in other words on string geometries supported by R-R three-form flux only. More generally, the $\AdS_3/\CFT_2$ correspondence is believed to hold for configurations involving NS5-branes and fundamental strings in addition to the D1- and D5-branes. In this case the string geometry is supported by a mixture of NS-NS and R-R three-form flux, giving a family of backgrounds related by Type IIB S-duality. The all-loop worldsheet S matrix for these backgrounds is known~\cite{Lloyd:2014bsa,Hoare:2013ida}, and we expect also these dualities to be governed by a quantum integrable system. It would be interesting to understand whether an integrable spin-chain picture emerges in this setting too. It would also be interesting to understand how recent work on higher-spin limits of $\AdS_3/\CFT_2$, such as~\cite{Gaberdiel:2014cha}, can be related to our findings. Recent progress on understanding the \emph{large} $\mathcal{N}=(4,4)$ $\AdS_3/\CFT_2$ duality~\cite{Tong:2014yna} opens up the possibility of establishing the expected connections to integrability more firmly.

The ``large-$N$'' method (which here means large $N_f$) is particularly useful in the study of the effective theory describing states localised near the origin of the Higgs branch, as was originally advocated by Witten~\cite{Witten:1997yu}. In this paper, we have used this approach to perform some simple computations of anomalous dimensions at leading order. But large-$N$ techniques have been very widely exploited for a long time and an extensive set of tools exists to perform calculations in this approach. As a result, we expect that these methods, combined with integrability, will lead to significant advances in understanding generic unprotected quantities in the $\AdS_3/\CFT_2$ correspondence in the near future.

\section*{Acknowledgements}
We would like to thank Gleb Arutyunov, Riccardo Borsato, Stefano Cremonesi, Nick Dorey, Sergey Frolov, Amihay Hanany, Chris Hull, Juan Maldacena, Joe Minahan, Sameer Murthy, Leonardo Rastelli, David Tong, Alessandro Torrielli, Arkady Tseytlin and Kostya Zarembo for helpful discussions. We are particularly grateful to Riccardo Borsato, Tom Lloyd and Alessandro Torrielli for collaborating with us on a number of related projects.
O.O.S.'s  work was supported by the ERC Advanced grant No.~290456,
 ``Gauge theory -- string theory duality''.
A.S.'s work is funded by the People Programme (Marie Curie Actions) of the European Union, Grant Agreement No.~317089 (GATIS). A.S. also acknowledges the hospitality at APCTP where part of this work was
done, as well as hospitality by {\'E}cole Normale Sup{\'e}rieure in Paris and interesting discussions with Benjamin Basso, Volodya Kazakov and Jan Troost.
B.S.\@ acknowledges funding support from an STFC Consolidated Grant  ``Theoretical Physics at City University''
ST/J00037X/1.

\appendix

\section{Gamma matrices}
\label{sec:gamma-matrices}

We use the gamma matrices
\begin{equation}
  \newcommand{\I}{\mathrlap{\,\mathds{1}}\hphantom{\sigma_a}}
  \begin{aligned}
    \Gamma^0 &= + i \sigma_2 \otimes \sigma_3 \otimes \sigma_3 , \\
    \Gamma^1 &= \phantom{i} {-} \sigma_1 \otimes \I \otimes \sigma_3 , \\
    \Gamma^2 &= \phantom{i} {-} \sigma_2 \otimes \sigma_2 \otimes \I , \\
    \Gamma^3 &= \phantom{i} {+} \sigma_2 \otimes \sigma_1 \otimes \sigma_3 , \\
    \Gamma^4 &= \phantom{i} {-} \sigma_1 \otimes \sigma_2 \otimes \sigma_1 , \\
    \Gamma^5 &= \phantom{i} {+} \sigma_1 \otimes \sigma_2 \otimes \sigma_2 .
  \end{aligned}
\end{equation}
In this basis a generic six-dimensional spinor can be written as a tri-spinor $\Psi_{\alpha_0 \alpha_1 \alpha_2}$.
Introducing the intertwiners
\begin{equation}
  C = -\Gamma^0 , \qquad
  B = \Gamma^{0125} , \qquad
  T = B^t C .
\end{equation}
the Gamma matrices satisfy
\begin{equation}
  C \Gamma^\mu C^{-1} = - ( \Gamma^m )^{\dag} , \qquad
  B \Gamma^\mu B^{-1} = - ( \Gamma^m )^{*} , \qquad
  T \Gamma^\mu T^{-1} = + ( \Gamma^m )^t .
\end{equation}
We also have
\begin{equation}
  \Gamma^{012345} = - \sigma_3 \otimes \mathds{1} \otimes \mathds{1}  , \qquad
  \Gamma^{01} = - \sigma_3 \otimes \sigma_3 \otimes \mathds{1} , \qquad
  \Gamma^{2345} = + \mathds{1} \otimes \sigma_3 \otimes \mathds{1} .
\end{equation}
It is also useful to note
\begin{equation}
  \newcommand{\I}{\mathrlap{\,\mathds{1}}\hphantom{\sigma_a}}
  \begin{aligned}
    C \Gamma^0 &= \I \otimes \I_4 , \\
    C \Gamma^1 &= \sigma_3 \otimes \hat{\gamma}^{2345} , \\
    C \Gamma^2 &= \I \otimes \hat{\gamma}^2 , \\
    C \Gamma^3 &= \I \otimes \hat{\gamma}^3 , \\
    C \Gamma^4 &= \sigma_3 \otimes \hat{\gamma}^4 , \\
    C \Gamma^5 &= \sigma_3 \otimes \hat{\gamma}^5 ,
  \end{aligned}
\end{equation}
where we have introduced the $\algSO(4)$ gamma matrices
\begin{equation}
  (\hat{\gamma}^i)_{a\dot{a} , b\dot{b}} = \begin{pmatrix} 0 & (\gamma^i)_{a\dot{b}} \\ (\tilde{\gamma}^i)_{\dot{a}b} & 0 \end{pmatrix} ,
\end{equation}
with
\begin{equation}
  \begin{aligned}
    \gamma^2 &= + \sigma_3 , \quad &
    \gamma^3 &= - i \mathds{1} , \quad &
    \gamma^4 &= + \sigma_2 , \quad &
    \gamma^5 &= + \sigma_1 , \quad &
    \tilde{\gamma}^i &= + ( \gamma^i )^\dag ,
  \end{aligned}
\end{equation}
and
\begin{equation}
  \hat{\gamma}^{2345} = \sigma_3 \otimes \mathds{1} .
\end{equation}

We also have
\begin{equation}
  \newcommand{\I}{\mathrlap{\,\mathds{1}}\hphantom{\sigma_a}}
  \begin{aligned}
    T \Gamma^0 &= + \I \otimes t , \\
    T \Gamma^1 &= + \I \otimes t \hat{\gamma}^{2345} , \\
    T \Gamma^2 &= + \I \otimes t \hat{\gamma}^2 , \\
    T \Gamma^3 &= + \I \otimes t \hat{\gamma}^3 , \\
    T \Gamma^4 &= + \sigma_3 \otimes t \hat{\gamma}^3 , \\
    T \Gamma^5 &= + \sigma_3 \otimes t \hat{\gamma}^5 , 
  \end{aligned}
\end{equation}
where the four-dimensional transpose intertwiner is given by
\begin{equation}
  t = \sigma_3 \otimes \epsilon .
\end{equation}

\section{\texorpdfstring{$G$}{G}-functions}
\label{app:G-functions}

It is useful to introduce the basic one-loop integrals
\begin{equation}
  \begin{aligned}
    \int \frac{d^D k}{(2\pi)^D} \frac{1}{k^{2\alpha} (k-p)^{2\beta}} &= \frac{1}{p^{2(\alpha+\beta-D/2)}} G(\alpha,\beta) , \\
    \int \frac{d^D k}{(2\pi)^D} \frac{k^{\mu}}{k^{2\alpha} (k-p)^{2\beta}} &= \frac{p^{\mu}}{p^{2(\alpha+\beta-D/2)}} G_1(\alpha,\beta) , \\
    \int \frac{d^D k}{(2\pi)^D} \frac{k\cdot(p-k)}{k^{2\alpha} (k-p)^{2\beta}} &= \frac{1}{p^{2(\alpha+\beta-1-D/2)}} G_2(\alpha,\beta) .
  \end{aligned}
\end{equation}
The function $G$ is given by
\begin{equation}
  G(\alpha,\beta) = \frac{1}{(4\pi)^{D/2}} \frac{\Gamma(\alpha+\beta-D/2) \Gamma(D/2-\alpha) \Gamma(D/2-\beta)}{\Gamma(\alpha)\Gamma(\beta)\Gamma(D-\alpha-\beta)} .
\end{equation}
The integrals with non-trivial numerators can be expressed in terms of the above integral as
\begin{equation}
  \begin{aligned}
    G_1(\alpha,\beta) &= \frac{1}{2} \bigl( G(\alpha,\beta) - G(\alpha,\beta-1) + G(\alpha-1,\beta) \bigr) , \\
    G_2(\alpha,\beta) &= \frac{1}{2} \bigl( G(\alpha,\beta) - G(\alpha,\beta-1) - G(\alpha-1,\beta) \bigr) .
  \end{aligned}
\end{equation}

\section{Regularisation of Feynman diagrams}
\label{app:feynman}

In this appendix we present the details of the Feynman diagram calculation discussed in section~\ref{sec:oneloop}.
Since we are in two dimensions the IR properties of the theory are rather delicate~\cite{Witten:1997yu}. What is more, the effective propagator for the vector multiplet fields that follows from the two-point function~\eqref{eq:phi-two-pt-fn} is logarithmic in momentum space. All this makes it difficult to evaluate the loop-momenta integrals. To overcome such technical complications we propose the following method. We will compute the one-loop anomalous dimensions in a large $N_f$ expansion for a general dimension $d>2$, following which we will take the limit $d\rightarrow 2$.\footnote{The loop-momentum integrals will be performed in dimensional regularisation---we will evaluate them first in dimension $D$ and then analyticaly continue $D\rightarrow d$.} This method is useful because the effective propagator for the scalars $\phi$  in dimension $d>2$ has a simple form in momentum space~\eqref{eq:phi-two-pt-fn-fourier}. It also allows us to retain much better control over the IR region of momentum integration. We believe this prescription is physically well motivated,  because we are interested in the anomalous dimensions of the $\CFT_2$ which are related to the UV properties of the theory. Furthermore, within the large-N approach anomalous dimensions are often computed in a similar manner for generic $d$~\cite{Lang:1991kp,Petkou:1994ad} and, when finite, the $d\rightarrow 2$ limit can be taken.

The leading order diagrams that we are interested in are given in equation~\eqref{eq:D-one-loop-diagrams}. The diagrams that give non-trivial flavour interactions are
\begin{equation}
  \label{eq:finitediagrams}
  \begin{tikzpicture}[baseline=-0.5ex]
    \coordinate (i1) at (0,-1.2cm);
    \coordinate (i2) at (1cm,-1.2cm);

    \coordinate (o1) at (0,1.2cm);
    \coordinate (o2) at (1cm,1.2cm);

    \coordinate (v1) at (0.5cm,-0.8cm);
    \coordinate (v2) at (0.5cm,+0.8cm);

    \draw [scalar,out=90,in=225] (i1) to (v1);
    \draw [scalar,out=90,in=315] (i2) to (v1);

    \draw [scalar,out=135,in=270] (v2) to (o1);
    \draw [scalar,out=45,in=270] (v2) to (o2);

    \draw [hyp scalar,out=45,in=315,looseness=1.5] (v1) to (v2);
    \draw [hyp scalar,out=135,in=225,looseness=1.5] (v1) to (v2);

    \fill [] (v1) circle [radius=0.0625cm];
    \fill [] (v2) circle [radius=0.0625cm];

    \draw [thick,rounded corners=2pt,fill=gray] ($(i1)+(-0.25cm,-0.125cm)$) -- ($(i1)+(-0.25cm,0)$) -- ($(i2)+(0.25cm,0)$) -- ($(i2)+(0.25cm,-0.125cm)$) -- cycle;
  \end{tikzpicture}
\qquad
  \begin{tikzpicture}[baseline=-0.5ex]
    \coordinate (i1) at (0,-1.2cm);
    \coordinate (i2) at (1cm,-1.2cm);

    \coordinate (o1) at (0,1.2cm);
    \coordinate (o2) at (1cm,1.2cm);

    \coordinate (v1) at (0.5cm,-0.8cm);
    \coordinate (v2) at (0.5cm,+0.8cm);

    \coordinate (v1b) at (0.5cm,-0.3cm);
    \coordinate (v2b) at (0.5cm,+0.3cm);

    \draw [scalar,out=90,in=225] (i1) to (v1);
    \draw [scalar,out=90,in=315] (i2) to (v1);

    \draw [scalar,out=135,in=270] (v2) to (o1);
    \draw [scalar,out=45,in=270] (v2) to (o2);

    \draw [hyp scalar,out=45,in=315,looseness=1.5] (v1) to (v1b);
    \draw [hyp scalar,out=135,in=225,looseness=1.5] (v1) to (v1b);

    \draw [hyp scalar,out=45,in=315,looseness=1.5] (v2b) to (v2);
    \draw [hyp scalar,out=135,in=225,looseness=1.5] (v2b) to (v2);

    \draw [gluon] (v1b) to (v2b);

    \fill [] (v1) circle [radius=0.0625cm];
    \fill [] (v2) circle [radius=0.0625cm];
    \fill [] (v1b) circle [radius=0.0625cm];
    \fill [] (v2b) circle [radius=0.0625cm];

    \draw [thick,rounded corners=2pt,fill=gray] ($(i1)+(-0.25cm,-0.125cm)$) -- ($(i1)+(-0.25cm,0)$) -- ($(i2)+(0.25cm,0)$) -- ($(i2)+(0.25cm,-0.125cm)$) -- cycle;
  \end{tikzpicture}
  \qquad
  \begin{tikzpicture}[baseline=-0.5ex]
    \coordinate (i1) at (0,-1.2cm);
    \coordinate (i2) at (1cm,-1.2cm);

    \coordinate (o1) at (0,1.2cm);
    \coordinate (o2) at (1cm,1.2cm);

    \coordinate (v1a) at (0.2cm,-0.8cm);
    \coordinate (v1c) at (0.8cm,-0.8cm);
    \coordinate (v2) at (0.5cm,+0.8cm);

    \coordinate (v1b) at (0.5cm,-0.3cm);
    \coordinate (v2b) at (0.5cm,+0.3cm);

    \draw [scalar,out=90,in=225] (i1) to (v1a);
    \draw [scalar,out=90,in=315] (i2) to (v1c);
    
    \draw [hyp fermion] (v1a) to (v1b);
    \draw [hyp fermion] (v1a) to (v1c);
    \draw [hyp fermion] (v1b) to (v1c);

    \draw [scalar,out=135,in=270] (v2) to (o1);
    \draw [scalar,out=45,in=270] (v2) to (o2);

    \draw [hyp scalar,out=45,in=315,looseness=1.5] (v2b) to (v2);
    \draw [hyp scalar,out=135,in=225,looseness=1.5] (v2b) to (v2);

    \draw [gluon] (v1b) to (v2b);

    \fill [] (v1a) circle [radius=0.0625cm];
    \fill [] (v1c) circle [radius=0.0625cm];
    \fill [] (v2) circle [radius=0.0625cm];
    \fill [] (v1b) circle [radius=0.0625cm];
    \fill [] (v2b) circle [radius=0.0625cm];

    \draw [thick,rounded corners=2pt,fill=gray] ($(i1)+(-0.25cm,-0.125cm)$) -- ($(i1)+(-0.25cm,0)$) -- ($(i2)+(0.25cm,0)$) -- ($(i2)+(0.25cm,-0.125cm)$) -- cycle;
  \end{tikzpicture}
  \qquad
  \begin{tikzpicture}[baseline=-0.5ex]
    \coordinate (i1) at (0,-1.2cm);
    \coordinate (i2) at (1cm,-1.2cm);

    \coordinate (o1) at (0,1.2cm);
    \coordinate (o2) at (1cm,1.2cm);

    \coordinate (v1) at (0.5cm,-0.8cm);
    \coordinate (v2a) at (0.2cm,+0.8cm);
    \coordinate (v2c) at (0.8cm,+0.8cm);

    \coordinate (v1b) at (0.5cm,-0.3cm);
    \coordinate (v2b) at (0.5cm,+0.3cm);

    \draw [hyp scalar,out=45,in=315,looseness=1.5] (v1) to (v1b);
    \draw [hyp scalar,out=135,in=225,looseness=1.5] (v1) to (v1b);

    \draw [scalar,out=90,in=225] (i1) to (v1);
    \draw [scalar,out=90,in=315] (i2) to (v1);

    \draw [hyp fermion] (v2a) to (v2b);
    \draw [hyp fermion] (v2a) to (v2c);
    \draw [hyp fermion] (v2b) to (v2c);

    \draw [scalar,out=135,in=270] (v2a) to (o1);
    \draw [scalar,out=45,in=270] (v2c) to (o2);
     
    \draw [gluon] (v1b) to (v2b);

    \fill [] (v1) circle [radius=0.0625cm];
    \fill [] (v2a) circle [radius=0.0625cm];
    \fill [] (v2c) circle [radius=0.0625cm];
    \fill [] (v1b) circle [radius=0.0625cm];
    \fill [] (v2b) circle [radius=0.0625cm];

    \draw [thick,rounded corners=2pt,fill=gray] ($(i1)+(-0.25cm,-0.125cm)$) -- ($(i1)+(-0.25cm,0)$) -- ($(i2)+(0.25cm,0)$) -- ($(i2)+(0.25cm,-0.125cm)$) -- cycle;
  \end{tikzpicture}
  \qquad
  \begin{tikzpicture}[baseline=-0.5ex]
    \coordinate (i1) at (0,-1.2cm);
    \coordinate (i2) at (1cm,-1.2cm);

    \coordinate (o1) at (0,1.2cm);
    \coordinate (o2) at (1cm,1.2cm);

    \coordinate (v1a) at (0.2cm,-0.8cm);
    \coordinate (v1c) at (0.8cm,-0.8cm);
    \coordinate (v2a) at (0.2cm,+0.8cm);
    \coordinate (v2c) at (0.8cm,+0.8cm);

    \coordinate (v1b) at (0.5cm,-0.3cm);
    \coordinate (v2b) at (0.5cm,+0.3cm);

    \draw [scalar,out=90,in=225] (i1) to (v1a);
    \draw [scalar,out=90,in=315] (i2) to (v1c);
    
    \draw [hyp fermion] (v1a) to (v1b);
    \draw [hyp fermion] (v1a) to (v1c);
    \draw [hyp fermion] (v1b) to (v1c);
    
    \draw [hyp fermion] (v2a) to (v2b);
    \draw [hyp fermion] (v2a) to (v2c);
    \draw [hyp fermion] (v2b) to (v2c);

    \draw [scalar,out=135,in=270] (v2a) to (o1);
    \draw [scalar,out=45,in=270] (v2c) to (o2);
     
    \draw [gluon] (v1b) to (v2b);

    \fill [] (v1a) circle [radius=0.0625cm];
    \fill [] (v1c) circle [radius=0.0625cm];
    \fill [] (v2a) circle [radius=0.0625cm];
    \fill [] (v2c) circle [radius=0.0625cm];
    \fill [] (v1b) circle [radius=0.0625cm];
    \fill [] (v2b) circle [radius=0.0625cm];

    \draw [thick,rounded corners=2pt,fill=gray] ($(i1)+(-0.25cm,-0.125cm)$) -- ($(i1)+(-0.25cm,0)$) -- ($(i2)+(0.25cm,0)$) -- ($(i2)+(0.25cm,-0.125cm)$) -- cycle;
  \end{tikzpicture},
\end{equation}
and
\begin{equation}
  \label{eq:divdiagram}
  \begin{tikzpicture}[baseline=-0.5ex]
    \coordinate (i1) at (0,-1.2cm);
    \coordinate (i2) at (1cm,-1.2cm);

    \coordinate (o1) at (0,1.2cm);
    \coordinate (o2) at (1cm,1.2cm);

    \coordinate (v1) at (0,-0.4cm);
    \coordinate (v2) at (0,+0.4cm);
    \coordinate (v3) at (1cm,-0.4cm);
    \coordinate (v4) at (1cm,+0.4cm);

    \draw [scalar] (i1) to (v1);
    \draw [scalar] (i2) to (v3);

    \draw [scalar] (v2) to (o1);
    \draw [scalar] (v4) to (o2);

    \draw [hyp fermion] (v2) to (v1) to (v3) to (v4) to (v2);

    \fill [] (v1) circle [radius=0.0625cm];
    \fill [] (v2) circle [radius=0.0625cm];
    \fill [] (v3) circle [radius=0.0625cm];
    \fill [] (v4) circle [radius=0.0625cm];

    \draw [thick,rounded corners=2pt,fill=gray] ($(i1)+(-0.25cm,-0.125cm)$) -- ($(i1)+(-0.25cm,0)$) -- ($(i2)+(0.25cm,0)$) -- ($(i2)+(0.25cm,-0.125cm)$) -- cycle;
  \end{tikzpicture}.
\end{equation}
The four diagrams in~\eqref{eq:finitediagrams} involving a gluon exchange all vanish by symmetry. As an example of this let us consider the lower subgraph of the second and fourth diagram containing a scalar bubble
\begin{equation}
  \begin{tikzpicture}[baseline=-0.525cm-0.5ex]
    \coordinate (i1) at (0,-1.2cm);
    \coordinate (i2) at (1cm,-1.2cm);

    \coordinate (o2) at (1cm,1.2cm);

    \coordinate (v1) at (0.5cm,-0.8cm);

    \coordinate (v2) at (0.5cm,-0.3cm);
    \coordinate (v3) at (0.5cm,+0.15cm);

    \draw [hyp scalar,out=45,in=315,looseness=1.5] (v1) to (v2);
    \draw [hyp scalar,out=135,in=225,looseness=1.5] (v1) to (v2);
    
    \draw [scalar,out=90,in=225] (i1) to (v1);
    \draw [scalar,out=90,in=315] (i2) to (v1);
    
    \draw [gluon] (v2) to (v3);

    \fill [] (v1) circle [radius=0.0625cm];
    \fill [] (v2) circle [radius=0.0625cm];

    \draw [thick,rounded corners=2pt,fill=gray] ($(i1)+(-0.25cm,-0.125cm)$) -- ($(i1)+(-0.25cm,0)$) -- ($(i2)+(0.25cm,0)$) -- ($(i2)+(0.25cm,-0.125cm)$) -- cycle;
  \end{tikzpicture}.
\end{equation}
This integral can be represented by the two diagrams
\begin{equation}
  \begin{tikzpicture}[baseline={([yshift=-0.5ex] v1)}]
    \coordinate (i) at (0,-2cm);

    \coordinate (v1) at (0cm,-1cm);
    \coordinate (v2) at (0cm,0cm);

    \coordinate (o) at (0cm,+0.3cm);

    \draw [thick,double] (i) .. controls ($(i)+(150:0.5cm)$) and ($(v1)+(210:0.5cm)$) ..  (v1);% node [pos=0.5,anchor=south,rotate=90] {\scriptsize $d/2-1$};
    \draw [thick,double] (i) .. controls ($(i)+( 30:0.5cm)$) and ($(v1)+(330:0.5cm)$) ..  (v1);% node [pos=0.5,anchor=north,rotate=90] {\scriptsize $d/2-1$};

    \draw [thick,derivative=0.5,postaction={decorate}] (v1) .. controls ($(v1)+(150:0.5cm)$) and ($(v2)+(210:0.5cm)$) ..  (v2) node [pos=0.5,anchor=east] {\scriptsize $\mu$};
    \draw [thick] (v2) .. controls ($(v2)+(330:0.5cm)$) and ($(v1)+( 30:0.5cm)$) ..  (v1);

    \draw [gluon] (v2) -- (o) node [anchor=west] {\scriptsize $\mu$};

    \fill [] (i) circle [radius=0.0625cm];
    \fill [] (v1) circle [radius=0.0625cm];
    \fill [] (v2) circle [radius=0.0625cm];

    \draw [-latex] ($(i)+(0,-0.55cm)$) -- ($(i)+(0,-0.15cm)$) node [pos=0,anchor=north] {\footnotesize $p$};
  \end{tikzpicture}
  \, + \,
  \begin{tikzpicture}[baseline={([yshift=-0.5ex] v1)}]
    \coordinate (i) at (0,-2cm);

    \coordinate (v1) at (0cm,-1cm);
    \coordinate (v2) at (0cm,0cm);

    \coordinate (o) at (0cm,+0.3cm);

    \draw [thick,double] (i) .. controls ($(i)+(150:0.5cm)$) and ($(v1)+(210:0.5cm)$) ..  (v1);% node [pos=0.5,anchor=south,rotate=90] {\scriptsize $d/2-1$};
    \draw [thick,double] (i) .. controls ($(i)+( 30:0.5cm)$) and ($(v1)+(330:0.5cm)$) ..  (v1);% node [pos=0.5,anchor=north,rotate=90] {\scriptsize $d/2-1$};

    \draw [thick] (v1) .. controls ($(v1)+(150:0.5cm)$) and ($(v2)+(210:0.5cm)$) ..  (v2);
    \draw [thick,derivative=0.5,postaction={decorate}] (v2) .. controls ($(v2)+(330:0.5cm)$) and ($(v1)+( 30:0.5cm)$) ..  (v1) node [pos=0.5,anchor=west] {\scriptsize $\mu$};

    \draw [gluon] (v2) -- (o) node [anchor=west] {\scriptsize $\mu$};

    \fill [] (i) circle [radius=0.0625cm];
    \fill [] (v1) circle [radius=0.0625cm];
    \fill [] (v2) circle [radius=0.0625cm];

    \draw [-latex] ($(i)+(0,-0.55cm)$) -- ($(i)+(0,-0.15cm)$) node [pos=0,anchor=north] {\footnotesize $p$};
  \end{tikzpicture},
\end{equation}
where the arrows indicate a factor of the corresponding momentum in the numerator of the integral. Reflecting the second diagram horizontally gives the first diagram but with the momentum insertion in the opposite direction so that the two diagrams sum to zero. A similar argument can be used to show that the fermion triangle appearing in the bottom of the third and fifth diagram of~\eqref{eq:finitediagrams} vanishes.

The remaining leftmost integral in~\eqref{eq:finitediagrams} is given by a product of two bubble integrals. One bubble consists of scalar fields from the vector multiplet  and one of scalars from the hypermultiplet, $I_{\text{bb}}=B_{\text{hyp}}B_{\text{vec}}$. Let us consider the bubble from the hypermultiplet. Power counting dictates that such an integral must be UV-convergent in any $d<4$, and in particular in $d=2$ where we want to compute it. Dimensional regularisation yields
\begin{equation}
    B_{\text{hyp}}=\int \frac{d^D k}{(2\pi)^D} \frac{1}{k^{2} (k-p)^{2}} = \frac{1}{p^{2(2-D/2)}} G(1,1)
    \approx \frac{1}{\pi p^2}\frac{1}{\epsilon},
\end{equation}
where in the last step we expanded around $D=2+\epsilon$, and the function $G$ was introduced in appendix~\ref{app:G-functions}. Clearly this pole is due to an~\emph{infrared} divergence, and as such will not contribute to the mixing matrix. The bubble of scalars from the vector multiplet on the other hand appears, by power-counting, to be polynomially UV divergent (but IR-finite). In dimensional regularisation we have
\begin{equation}
    B_{\text{vec}}=\int \frac{d^D k}{(2\pi)^D} \frac{1}{k^{2(d/2-1)} (k-p)^{2(d/2-1)}} = \frac{1}{p^{2(d-2-D/2)}} G\big(\frac{d}{2}-1,\frac{d}{2}-1\big).
\end{equation}
We want to see whether this expression is UV divergent in dimension~$d$, \ie, we set~$D=d+\epsilon$. We obtain, up to numerical prefactors
\begin{equation}
    B_{\text{vec}}=p^{2(2-d/2)}\frac{1}{(d/2-2)^2\Gamma[d/2-2]}+O(\epsilon)\,.
\end{equation}
Not only is this expression regular at $d=2$, but in fact it \emph{vanishes}, and it does so regardless of how we send $D\to d\to2$.%
\footnote{%
Interestingly, it vanishes even in $d=0$, where we would expect the $d=2$ quadratic divergence to manifest itself. This is in a way similar to what happens to the renormalisation of the photon two-point function in quantum electrodynamics.
} 
Clearly the bubbles are well-behaved in the UV, and~$I_{\text{bb}}$ cannot contribute to the mixing matrix.

We are left with the fermion box diagram~\eqref{eq:divdiagram}. This diagrams gives rise to two integrals $I_1$ and $I_2$ which differ in the chiralities of the fermions. Diagrammatically these integrals can be represented by
\begin{equation}
  I_1 =\!\!\!\!
  \begin{tikzpicture}[baseline={([yshift=-0.5ex] 0,+1.25cm)}]
    \coordinate (i) at (0,0);

    \coordinate (o1) at (-0.75cm,+2.5cm);
    \coordinate (o2) at (+0.75cm,+2.5cm);

    \coordinate (v1) at (-0.75cm,+1cm);
    \coordinate (v2) at (-0.75cm,+2cm);
    \coordinate (v3) at (+0.75cm,+1cm);
    \coordinate (v4) at (+0.75cm,+2cm);

    \draw [thick,double] (i) .. controls ($(i)+(150:0.5cm)$) and ($(v1)+(270:0.5cm)$) ..  (v1) node [pos=0.5,anchor=north,rotate=315] {\scriptsize $d/2-1$};
    \draw [thick,double] (i) .. controls ($(i)+(30:0.5cm)$) and ($(v3)+(270:0.5cm)$) ..  (v3)  node [pos=0.5,anchor=north,rotate=45] {\scriptsize $d/2-1$};

    \draw [thick,derivative=0.5,postaction={decorate}] (v1) -- (v2) node [pos=0.5,anchor=east] {\footnotesize $+$};
    \draw [thick,derivative=0.5,postaction={decorate}] (v2) -- (v4) node [pos=0.5,anchor=south] {\footnotesize $-$};
    \draw [thick,derivative=0.5,postaction={decorate}] (v4) -- (v3) node [pos=0.5,anchor=west] {\footnotesize $+$};
    \draw [thick,derivative=0.5,postaction={decorate}] (v3) -- (v1) node [pos=0.5,anchor=north] {\footnotesize $-$};

    \draw [thick,double] (v2) -- (o1);
    \draw [thick,double] (v4) -- (o2);

    \fill [] (i) circle [radius=0.0625cm];
    \fill [] (v1) circle [radius=0.0625cm];
    \fill [] (v2) circle [radius=0.0625cm];
    \fill [] (v3) circle [radius=0.0625cm];
    \fill [] (v4) circle [radius=0.0625cm];

    \draw [-latex] ($(v2)+(-0.2cm,+0.05cm)$) -- ($(o1)+(-0.2cm,-0.05cm)$) node [pos=0.5,anchor=east] {\footnotesize $p_1$};
    \draw [-latex] ($(v4)+(+0.2cm,+0.05cm)$) -- ($(o2)+(+0.2cm,-0.05cm)$) node [pos=0.5,anchor=west] {\footnotesize $p_2$};

    \draw [-latex] ($(i)+(0,-0.55cm)$) -- ($(i)+(0,-0.15cm)$) node [pos=0,anchor=north] {\footnotesize $p_1+p_2$};

  \end{tikzpicture}
  \qquad
  I_2 = \!\!\!\!
  \begin{tikzpicture}[baseline={([yshift=-0.5ex] 0,+1.25cm)}]
    \coordinate (i) at (0,0);

    \coordinate (o1) at (-0.75cm,+2.5cm);
    \coordinate (o2) at (+0.75cm,+2.5cm);

    \coordinate (v1) at (-0.75cm,+1cm);
    \coordinate (v2) at (-0.75cm,+2cm);
    \coordinate (v3) at (+0.75cm,+1cm);
    \coordinate (v4) at (+0.75cm,+2cm);

    \draw [thick,double] (i) .. controls ($(i)+(150:0.5cm)$) and ($(v1)+(270:0.5cm)$) ..  (v1) node [pos=0.5,anchor=north,rotate=315] {\scriptsize $d/2-1$};
    \draw [thick,double] (i) .. controls ($(i)+(30:0.5cm)$) and ($(v3)+(270:0.5cm)$) ..  (v3)  node [pos=0.5,anchor=north,rotate=45] {\scriptsize $d/2-1$};

    \draw [thick,derivative=0.5,postaction={decorate}] (v1) -- (v2) node [pos=0.5,anchor=east] {\footnotesize $-$};
    \draw [thick,derivative=0.5,postaction={decorate}] (v2) -- (v4) node [pos=0.5,anchor=south] {\footnotesize $+$};
    \draw [thick,derivative=0.5,postaction={decorate}] (v4) -- (v3) node [pos=0.5,anchor=west] {\footnotesize $-$};
    \draw [thick,derivative=0.5,postaction={decorate}] (v3) -- (v1) node [pos=0.5,anchor=north] {\footnotesize $+$};

    \draw [thick,double] (v2) -- (o1);
    \draw [thick,double] (v4) -- (o2);

    \fill [] (i) circle [radius=0.0625cm];
    \fill [] (v1) circle [radius=0.0625cm];
    \fill [] (v2) circle [radius=0.0625cm];
    \fill [] (v3) circle [radius=0.0625cm];
    \fill [] (v4) circle [radius=0.0625cm];

    \draw [-latex] ($(v2)+(-0.2cm,+0.05cm)$) -- ($(o1)+(-0.2cm,-0.05cm)$) node [pos=0.5,anchor=east] {\footnotesize $p_1$};
    \draw [-latex] ($(v4)+(+0.2cm,+0.05cm)$) -- ($(o2)+(+0.2cm,-0.05cm)$) node [pos=0.5,anchor=west] {\footnotesize $p_2$};

    \draw [-latex] ($(i)+(0,-0.55cm)$) -- ($(i)+(0,-0.15cm)$) node [pos=0,anchor=north] {\footnotesize $p_1+p_2$};

  \end{tikzpicture}
\end{equation}
The arrows on the propagators in the boxes represent a factor of momentum in the numerator, with the $\pm$ next to the arrow indicating the component. The label below the double lines at the bottom serve as a reminder of the non-canonical weight of the propagator of the scalar in the vector multiplet. To evaluate the UV divergence of these integrals while avoiding IR divergences we can set the external momenta to $p_1 = -p_2 = p$.

We will start by showing that the two integrals $I_1$ and $I_2$ have the same UV divergence. Let us therefore consider the difference between these integrals
\begin{equation}
  I_1 - I_2 =
  \begin{tikzpicture}[baseline={([yshift=-.5ex] current bounding box.center)}]
    \useasboundingbox (-1.375cm,+0.75cm) rectangle (+1.375cm,-2.0cm);

    \coordinate (i) at (-1.25cm,0);
    \coordinate (o) at (+1.25cm,0);

    \coordinate (v1) at (-0.75cm,0);
    \coordinate (v2) at (+0.75cm,0);

    \coordinate (v3) at (-0.75cm,-1.25cm);
    \coordinate (v4) at (+0.75cm,-1.25cm);

    \draw [thick,double] (i) -- (v1);
    \draw [thick,double] (v2) -- (o);

    \draw [thick,derivative=0.5,postaction={decorate}] (v3) -- (v1) node [pos=0.5,anchor=east] {\footnotesize $+$};
    \draw [thick,derivative=0.5,postaction={decorate}] (v2) -- (v4) node [pos=0.5,anchor=west] {\footnotesize $+$};

    \draw [thick,derivative=0.5,postaction={decorate}] 
    (v1) .. controls ($(v1)+(45:0.75cm)$) and ($(v2)+(135:0.75cm)$) .. (v2) node [pos=0.5,anchor=south] {\footnotesize $-$};

    \draw [thick,derivative=0.5,postaction={decorate}]
    (v4) .. controls ($(v4)+(135:0.75cm)$) and ($(v3)+(45:0.75cm)$) .. (v3) node [pos=0.5,anchor=north] {\footnotesize $-$};

    \draw [thick] (v4) .. controls ($(v4)+(225:0.75cm)$) and ($(v3)+(315:0.75cm)$) .. (v3) node [pos=0.5,anchor=north] {\scriptsize $d-2$};

    \fill [] (v1) circle [radius=0.0625cm];
    \fill [] (v2) circle [radius=0.0625cm];
    \fill [] (v3) circle [radius=0.0625cm];
    \fill [] (v4) circle [radius=0.0625cm];

  \end{tikzpicture}
  -
  \begin{tikzpicture}[baseline={([yshift=-.5ex] current bounding box.center)}]
    \useasboundingbox (-1.375cm,+0.25cm) rectangle (+1.375cm,-1.5cm);

    \coordinate (i) at (-1.25cm,0);
    \coordinate (o) at (+1.25cm,0);

    \coordinate (v1) at (-0.75cm,0);
    \coordinate (v2) at (+0.75cm,0);

    \coordinate (v3) at (-0.75cm,-1.25cm);
    \coordinate (v4) at (+0.75cm,-1.25cm);

    \draw [thick,double] (i) -- (v1);
    \draw [thick,double] (v2) -- (o);

    \draw [thick,derivative=0.5,postaction={decorate}] (v3) -- (v1) node [pos=0.5,anchor=east] {\footnotesize $-$};
    \draw [thick,derivative=0.5,postaction={decorate}] (v2) -- (v4) node [pos=0.5,anchor=west] {\footnotesize $-$};

    \draw [thick,derivative=0.5,postaction={decorate}] 
    (v1) .. controls ($(v1)+(45:0.75cm)$) and ($(v2)+(135:0.75cm)$) .. (v2) node [pos=0.5,anchor=south] {\footnotesize $+$};

    \draw [thick,derivative=0.5,postaction={decorate}]
    (v4) .. controls ($(v4)+(135:0.75cm)$) and ($(v3)+(45:0.75cm)$) .. (v3) node [pos=0.5,anchor=north] {\footnotesize $+$};

    \draw [thick] (v4) .. controls ($(v4)+(225:0.75cm)$) and ($(v3)+(315:0.75cm)$) .. (v3) node [pos=0.5,anchor=north] {\scriptsize $d-2$};

    \fill [] (v1) circle [radius=0.0625cm];
    \fill [] (v2) circle [radius=0.0625cm];
    \fill [] (v3) circle [radius=0.0625cm];
    \fill [] (v4) circle [radius=0.0625cm];

  \end{tikzpicture}
\end{equation}
Since we have set $p_1 + p_2 = 0$, the two scalar propagators at the bottom have merged into a single propagator of weight $d-2$. We can now perform the bubble integrals to obtain
\begin{equation}
  I_1 - I_2 =
  G_1(1,d-2) \Biggl(
  \begin{tikzpicture}[baseline={([yshift=-.5ex] 0,0)}]
    \coordinate (i) at (-1.25cm,0);
    \coordinate (o) at (+1.25cm,0);

    \coordinate (v1) at (-0.75cm,0);
    \coordinate (v2) at (+0.75cm,0);

    \draw [thick,double] (i) -- (v1);
    \draw [thick,double] (v2) -- (o);

    \draw [thick,derivative=0.5,postaction={decorate}] 
    (v1) .. controls ($(v1)+(45:0.75cm)$) and ($(v2)+(135:0.75cm)$) .. (v2) node [pos=0.5,anchor=south] {\footnotesize $-$};

    \draw [thick]
    (v2) .. controls ($(v2)+(225:0.75cm)$) and ($(v1)+(315:0.75cm)$) .. (v1) node [pos=0.5,anchor=north] {\scriptsize $d-\frac{D}{2}+1$};

    \draw [thick,derivative=0.5,decorate] 
    (v2) .. controls ($(v2)+(225:0.75cm)$) and ($(v1)+(315:0.75cm)$) .. (v1) node [pos=0.5,anchor=south] {\footnotesize $-$};

    \draw [thick,derivative=0.25,decorate] 
    (v2) .. controls ($(v2)+(225:0.75cm)$) and ($(v1)+(315:0.75cm)$) .. (v1) node [pos=0.25,anchor=south] {\footnotesize $+$};

    \draw [thick,derivative=0.75,decorate] 
    (v2) .. controls ($(v2)+(225:0.75cm)$) and ($(v1)+(315:0.75cm)$) .. (v1) node [pos=0.75,anchor=south] {\footnotesize $+$};

    \fill [] (v1) circle [radius=0.0625cm];
    \fill [] (v2) circle [radius=0.0625cm];
  \end{tikzpicture}
  -
  \begin{tikzpicture}[baseline={([yshift=-.5ex] 0,0)}]
    \coordinate (i) at (-1.25cm,0);
    \coordinate (o) at (+1.25cm,0);

    \coordinate (v1) at (-0.75cm,0);
    \coordinate (v2) at (+0.75cm,0);

    \draw [thick,double] (i) -- (v1);
    \draw [thick,double] (v2) -- (o);

    \draw [thick,derivative=0.5,postaction={decorate}] 
    (v1) .. controls ($(v1)+(45:0.75cm)$) and ($(v2)+(135:0.75cm)$) .. (v2) node [pos=0.5,anchor=south] {\footnotesize $+$};

    \draw [thick]
    (v2) .. controls ($(v2)+(225:0.75cm)$) and ($(v1)+(315:0.75cm)$) .. (v1) node [pos=0.5,anchor=north] {\scriptsize $d-\frac{D}{2}+1$};

    \draw [thick,derivative=0.5,decorate] 
    (v2) .. controls ($(v2)+(225:0.75cm)$) and ($(v1)+(315:0.75cm)$) .. (v1) node [pos=0.5,anchor=south] {\footnotesize $+$};

    \draw [thick,derivative=0.25,decorate] 
    (v2) .. controls ($(v2)+(225:0.75cm)$) and ($(v1)+(315:0.75cm)$) .. (v1) node [pos=0.25,anchor=south] {\footnotesize $-$};

    \draw [thick,derivative=0.75,decorate] 
    (v2) .. controls ($(v2)+(225:0.75cm)$) and ($(v1)+(315:0.75cm)$) .. (v1) node [pos=0.75,anchor=south] {\footnotesize $-$};

    \fill [] (v1) circle [radius=0.0625cm];
    \fill [] (v2) circle [radius=0.0625cm];
  \end{tikzpicture}
  \Biggr) .
\end{equation}
The remaining loop integrals are tensor valued and anti-symmetrised. Since the only available tensors are the metric and the external momentum $p$, the difference of the two integrals vanishes and the UV divergences of $I_1$ and $I_2$ coincide.\footnote{%
  Note that the full integrals $I_1$ and $I_2$ are not necessarily identical. However, because the UV divergence can be evaluated at $p_1=-p_2$ the difference of the full integrals is finite.%
} %

Since the integrals $I_1$ and $I_2$ have the same UV divergence we can consider their sum, which can be interpreted as a single fermion box diagram with a Dirac fermion running in the loop. This leads to an integral of the form
\begin{equation}
  I_1 + I_2 = \int \frac{d^D k \, d^D q}{(2\pi)^{2D}} \frac{\tr \bigl( \slashed{k} (\slashed{k}-\slashed{p}_1) (\slashed{k}+\slashed{p}_2) (\slashed{k}-\slashed{q})\bigr)}{k^2 (k-p_1)^2 (k+p_2)^2 (k-q)^2 (q-p_1)^{2(d/2-1)} (q+p_2)^{2(d/2-1)}}.
\end{equation}
Again the UV divergence can be evaluated by setting $p_1 = -p_2 = p$.
The trace in the numerator gives rise to three different contractions, which diagrammatically can be represented as
\begin{equation}
  I_1 + I_2 =
  4\Biggl(
  \begin{tikzpicture}[baseline={([yshift=-.5ex] current bounding box.center)}]
    \useasboundingbox (-1.375cm,+0.75cm) rectangle (+1.375cm,-2.0cm);

    \coordinate (i) at (-1.25cm,0);
    \coordinate (o) at (+1.25cm,0);

    \coordinate (v1) at (-0.75cm,0);
    \coordinate (v2) at (+0.75cm,0);

    \coordinate (v3) at (-0.75cm,-1.25cm);
    \coordinate (v4) at (+0.75cm,-1.25cm);

    \draw [thick,double] (i) -- (v1);
    \draw [thick,double] (v2) -- (o);

    \draw [thick,derivative=0.5,postaction={decorate}] (v3) -- (v1) node [pos=0.5,anchor=east] {\footnotesize $\mu$};
    \draw [thick,derivative=0.5,postaction={decorate}] (v2) -- (v4) node [pos=0.5,anchor=west] {\footnotesize $\nu$};

    \draw [thick,derivative=0.5,postaction={decorate}] 
    (v1) .. controls ($(v1)+(45:0.75cm)$) and ($(v2)+(135:0.75cm)$) .. (v2) node [pos=0.5,anchor=south] {\footnotesize $\mu$};

    \draw [thick,derivative=0.5,postaction={decorate}]
    (v4) .. controls ($(v4)+(135:0.75cm)$) and ($(v3)+(45:0.75cm)$) .. (v3) node [pos=0.5,anchor=north] {\footnotesize $\nu$};

    \draw [thick] (v4) .. controls ($(v4)+(225:0.75cm)$) and ($(v3)+(315:0.75cm)$) .. (v3) node [pos=0.5,anchor=north] {\scriptsize $d-2$};

    \fill [] (v1) circle [radius=0.0625cm];
    \fill [] (v2) circle [radius=0.0625cm];
    \fill [] (v3) circle [radius=0.0625cm];
    \fill [] (v4) circle [radius=0.0625cm];

  \end{tikzpicture}
  +
  \begin{tikzpicture}[baseline={([yshift=-.5ex] current bounding box.center)}]
    \useasboundingbox (-1.375cm,+0.25cm) rectangle (+1.375cm,-1.5cm);

    \coordinate (i) at (-1.25cm,0);
    \coordinate (o) at (+1.25cm,0);

    \coordinate (v1) at (-0.75cm,0);
    \coordinate (v2) at (+0.75cm,0);

    \coordinate (v3) at (-0.75cm,-1.25cm);
    \coordinate (v4) at (+0.75cm,-1.25cm);

    \draw [thick,double] (i) -- (v1);
    \draw [thick,double] (v2) -- (o);

    \draw [thick,derivative=0.5,postaction={decorate}] (v3) -- (v1) node [pos=0.5,anchor=east] {\footnotesize $\nu$};
    \draw [thick,derivative=0.5,postaction={decorate}] (v2) -- (v4) node [pos=0.5,anchor=west] {\footnotesize $\mu$};

    \draw [thick,derivative=0.5,postaction={decorate}] 
    (v1) .. controls ($(v1)+(45:0.75cm)$) and ($(v2)+(135:0.75cm)$) .. (v2) node [pos=0.5,anchor=south] {\footnotesize $\mu$};

    \draw [thick,derivative=0.5,postaction={decorate}]
    (v4) .. controls ($(v4)+(135:0.75cm)$) and ($(v3)+(45:0.75cm)$) .. (v3) node [pos=0.5,anchor=north] {\footnotesize $\nu$};

    \draw [thick] (v4) .. controls ($(v4)+(225:0.75cm)$) and ($(v3)+(315:0.75cm)$) .. (v3) node [pos=0.5,anchor=north] {\scriptsize $d-2$};

    \fill [] (v1) circle [radius=0.0625cm];
    \fill [] (v2) circle [radius=0.0625cm];
    \fill [] (v3) circle [radius=0.0625cm];
    \fill [] (v4) circle [radius=0.0625cm];

  \end{tikzpicture}
  -
  \begin{tikzpicture}[baseline={([yshift=-.5ex] current bounding box.center)}]
    \useasboundingbox (-1.375cm,+0.25cm) rectangle (+1.375cm,-1.5cm);

    \coordinate (i) at (-1.25cm,0);
    \coordinate (o) at (+1.25cm,0);

    \coordinate (v1) at (-0.75cm,0);
    \coordinate (v2) at (+0.75cm,0);

    \coordinate (v3) at (-0.75cm,-1.25cm);
    \coordinate (v4) at (+0.75cm,-1.25cm);

    \draw [thick,double] (i) -- (v1);
    \draw [thick,double] (v2) -- (o);

    \draw [thick,derivative=0.5,postaction={decorate}] (v3) -- (v1) node [pos=0.5,anchor=east] {\footnotesize $\nu$};
    \draw [thick,derivative=0.5,postaction={decorate}] (v2) -- (v4) node [pos=0.5,anchor=west] {\footnotesize $\nu$};

    \draw [thick,derivative=0.5,postaction={decorate}] 
    (v1) .. controls ($(v1)+(45:0.75cm)$) and ($(v2)+(135:0.75cm)$) .. (v2) node [pos=0.5,anchor=south] {\footnotesize $\mu$};

    \draw [thick,derivative=0.5,postaction={decorate}]
    (v4) .. controls ($(v4)+(135:0.75cm)$) and ($(v3)+(45:0.75cm)$) .. (v3) node [pos=0.5,anchor=north] {\footnotesize $\mu$};

    \draw [thick] (v4) .. controls ($(v4)+(225:0.75cm)$) and ($(v3)+(315:0.75cm)$) .. (v3) node [pos=0.5,anchor=north] {\scriptsize $d-2$};

    \fill [] (v1) circle [radius=0.0625cm];
    \fill [] (v2) circle [radius=0.0625cm];
    \fill [] (v3) circle [radius=0.0625cm];
    \fill [] (v4) circle [radius=0.0625cm];
  \end{tikzpicture}
  \Biggr).
\end{equation}
In the above diagrams the arrows again indicate factors of momenta in the numerator, and the indices $\mu$ and $\nu$ indicate which momenta are to be contracted. Performing the one-loop bubble integral we are left with three identical terms which add up to
\begin{equation}
  \begin{aligned}
    I_1 + I_2 &=
    4 \, G_1(1,d-2) \,
    \begin{tikzpicture}[baseline={([yshift=-.5ex] 0,0)}]
      \coordinate (i) at (-1.25cm,0);
      \coordinate (o) at (+1.25cm,0);
      
      \coordinate (v1) at (-0.75cm,0);
      \coordinate (v2) at (+0.75cm,0);

      \draw [thick,double] (i) -- (v1);
      \draw [thick,double] (v2) -- (o);

      \draw [thick,derivative=0.5,postaction={decorate}] 
      (v1) .. controls ($(v1)+(45:0.75cm)$) and ($(v2)+(135:0.75cm)$) .. (v2) node [pos=0.5,anchor=south] {\footnotesize $\mu$};

      \draw [thick]
      (v2) .. controls ($(v2)+(225:0.75cm)$) and ($(v1)+(315:0.75cm)$) .. (v1) node [pos=0.5,anchor=north] {\scriptsize $d-\frac{D}{2}$};

      \draw [thick,derivative=0.5,decorate] 
      (v2) .. controls ($(v2)+(225:0.75cm)$) and ($(v1)+(315:0.75cm)$) .. (v1) node [pos=0.5,anchor=south] {\footnotesize $\mu$};

      \fill [] (v1) circle [radius=0.0625cm];
      \fill [] (v2) circle [radius=0.0625cm];
    \end{tikzpicture}
    \\
    &=
    4 \, G_1(1,d-2)
    \int \frac{d^D k}{(2\pi)^D} \frac{k \cdot (p-k)}{k^2 (k-p)^{2(d-D/2)}} .
  \end{aligned}
\end{equation}
If we go ahead and perform the remaining integration by dimensional regularisation we find
\begin{equation}
  I_1 + I_2 = -\frac{4}{p^{2(d-D)}}G_1(1,d-2) G_2(1,d-D/2) 
\end{equation}
which has a simple pole at $d=D$ for any $d>2$. However, taking the limit $d\to2$ in this expression is subtle.
Instead, we re-express the integral using Feynman parameters as
\begin{equation}\label{eq:fermion-box-feynman-parameters}
  \begin{aligned}
    I_1 + I_2 = 
    - &4 \, G_1(1,d-2) \int \frac{d^D k}{(2\pi)^D} \frac{1}{k^{2(d-\frac{D}{2})}} \\
    + &4 (d-\tfrac{D}{2}) G_1(1,d-2) \int \frac{d^D k}{(2\pi)^D} \int_0^1 dx \,
    \frac{x^{d-\frac{D}{2}-1}  ( p \cdot k + x p^2 )}{( k^2 + x(1-x) p^2 )^{d-\frac{D}{2}+1} } .
  \end{aligned}
\end{equation}
The integral in the second line is UV convergent. Performing the integral in the first line we get
\begin{equation}
  I_1 + I_2 \approx
  -\frac{8G_1(1,d-2)}{(4\pi)^{D/2} \Gamma(D/2)} \int_{\mu}^{\Lambda} \frac{dk}{k^{2(d-D)+1}}
  = - \frac{4G_1(1,d-2)}{(4\pi)^{D/2} \Gamma(D/2)} \frac{\Lambda^{2(D-d)} - \mu^{2(D-d)}}{D - d} ,
\end{equation}
where we have dropped terms that are not UV divergent and $\Lambda$ and $\mu$ are cutoffs at large and small momentum.
Using the definition of $G_1$ we get
\begin{equation}
  I_1 + I_2 \approx \frac{4\pi}{(4\pi)^D} \frac{\csc((d-D/2)\pi)}{\Gamma(d-2) \Gamma(D-d+2)} \frac{\Lambda^{2(D-d)} - \mu^{2(D-d)}}{d-D}\, .
\end{equation}
In the above calculation the dimension $d$ has been introduced in the propagator to avoid IR divergences, while $D$ has been introduced in the integrations to avoid UV divergences. Hence we now need to take the limit where $d$ and $D$ approach $2$ with
\begin{equation}
  2 < D < d .
\end{equation}
A natural way to do this is to first send $D \to d$, and then $d \to 2$, which results in
\begin{equation}
  I_1 + I_2 \approx \frac{8\pi}{(4\pi)^d} \frac{\csc(d\pi/2)}{\Gamma(d-2)} \log \frac{\Lambda}{\mu} \approx -\frac{1}{\pi^2} \log\frac{\Lambda}{\mu} ,
\end{equation}
so that the fermionic box diagram~\eqref{eq:divdiagram} is UV divergent. This expression also appears IR divergent when we send $\mu$ to zero. However, in the full fermion box integral with the momentum assignment we have made there is no such divergence, so this IR divergence is cancelled by a contribution from the UV convergent integral in~\eqref{eq:fermion-box-feynman-parameters} that we have dropped.

\bibliographystyle{nb}
\bibliography{higgs-spin-chain}

\end{document}